\documentclass[11pt]{article}

\usepackage{amsmath}
\usepackage{amssymb}

\usepackage{graphicx}

\usepackage{cite}
\usepackage{color} 

\usepackage[labelfont=bf,labelsep=period,justification=raggedright]{caption}

\makeatletter
\renewcommand{\@biblabel}[1]{\quad#1.}
\makeatother

\date{}

\usepackage{comment}

\def\azul{\textcolor{blue}}

\def\rojo{\textcolor{red}}

\newcommand{\mylab}[3]{\raisebox{#2}[0mm][0mm]{%
\makebox[0mm][l]{\hspace*{#1}#3}}}
\def\spacce#1{\hskip #1pt}
\def\drawline#1#2{\raise 2.5pt\vbox{\hrule width #1pt height #2pt}}
\def\solid{\drawline{24}{.5}\nobreak}

\def\bdash{\hbox{\drawline{7}{.5}\spacce{2}}}

\def\dashed{\bdash\bdash\bdash\nobreak}

\def\chndot{\hbox%
{\drawline{9.5}{.5}\spacce{2}\drawline{1}{.5}\spacce{2}\drawline{9.5}{.5}}\nobreak }

\def\trian{\raise 1.25pt\hbox{$\scriptstyle\triangle$}\nobreak}
\def\leftrian{\raise 0.75pt\hbox{$\displaystyle\triangleleft$}\nobreak}
\def\rightrian{\raise 0.75pt\hbox{$\displaystyle\triangleright$}\nobreak}

\def\dtrian{\raise 1.25pt\hbox%
{$\scriptscriptstyle\bigtriangledown$}\nobreak}

\def\squar{\raise 1.25pt\hbox{$\scriptstyle\Box$}\nobreak}

\def\diamon{\raise 1.25pt\hbox{$\scriptstyle\diamond$}\nobreak}

\def\solidtrian{$\blacktriangle$\nobreak}

\def\linesolidtrian{\hbox%
{\drawline{8}{.5}\spacce{2}\solidtrian\drawline{8}{.5}}\nobreak}
\def\solidsquar{$\blacksquare$\nobreak}
\def\linesolidsquar{\hbox%
{\drawline{8}{.5}\spacce{2}\solidsquar\drawline{8}{.5}}\nobreak}
\def\solidcircle{$\bullet$\nobreak}
\def\linesolidcircle{\hbox%
{\drawline{8}{.5}\spacce{2}\solidcircle\spacce{2}\drawline{8}{.5}}\nobreak}

\def\dd{{\, \rm{d}}}

\def\beq{\begin{equation}}
\def\eeq{\end{equation}}

\def\abol{\textbf{\emph{(a)}}}
\def\bbol{\textbf{\emph{(b)}}}
\def\cbol{\textbf{\emph{(c)}}}
\def\dbol{\textbf{\emph{(d)}}}
\def\ebol{\textbf{\emph{(e)}}}
\def\fbol{\textbf{\emph{(f)}}}
\def\gbol{\textbf{\emph{(g)}}}

\def\aaa{{\it a}}
\def\bbb{{\it b}}
\def\ccc{{\it c}}
\def\ddd{{\it d}}
\def\eee{{\it e}}
\def\fff{{\it f}}
\def\ggg{{\it g}}
\def\hhh{{\it h}}
\def\iii{{\it i}}
\def\jjj{{\it j}}
\def\kkk{{\it k}}
\def\lll{{\it l}}
\def\citalajim03{Del \'Alamo \& Jim\'enez (2003)}
\newcommand\ie{i.e.\ }
\newcommand\eg{e.g.\ }
\newcommand\etal{{\it et al.}}

\begin{document}

%
\begin{flushleft}
{\Large \textbf{Three-Dimensional Quantification of Cellular Traction Forces
and Mechanosensing of Thin Substrata by Fourier Traction Force Microscopy} }
%
\\ 
Juan C. del \'Alamo$^{1,2,\ast}$, 
Ruedi Meili$^{1,3}$, 
Bego\~na \'Alvarez-Gonz\'alez$^{1}$,
Baldomero Alonso-Latorre$^{1}$,
Effie Bastounis$^{1,3,4}$, 
Richard Firtel$^{3}$,
Juan C. Lasheras$^{1,2,4}$,
\\
\bf{1} Mechanical and Aerospace Engineering Department, University of
California San Diego, La Jolla, CA, USA.
\\
\bf{2} Institute for Engineering in Medicine, University of California San
Diego, La Jolla, CA, USA.
\\
\bf{3} Division of Cell and Developmental Biology, University of California San
Diego, La Jolla, CA, USA.
\\
\bf{4} Bioengineering Department, University of California San Diego, La Jolla,
CA, USA.
\\
$\ast$ E-mail: jalamo@ucsd.edu
\end{flushleft}

\section*{Abstract}
We introduce a novel three-dimensional (3D) traction force microscopy (TFM)
method motivated by the recent discovery that cells adhering on plane surfaces
exert both in-plane and out-of-plane traction stresses. 
We measure the 3D deformation of the substratum on a thin layer near its
surface, and input this information into an exact analytical solution of the
elastic equilibrium equation.
These operations are performed in the Fourier domain with high computational
efficiency, allowing to obtain the 3D traction stresses from raw microscopy
images virtually in real time.
We also characterize the error of previous two-dimensional (2D) TFM methods
that neglect the out-of-plane component of the traction stresses.
This analysis reveals that, under certain combinations of experimental
parameters (\ie cell size, substratums' thickness and Poisson's ratio), the
accuracy of 2D TFM methods is minimally affected by neglecting the out-of-plane
component of the traction stresses.
Finally, we consider the cell's mechanosensing of substratum thickness by 3D
traction stresses, finding that,
when cells adhere on thin substrata, their out-of-plane traction stresses can
reach four times deeper into the substratum than their in-plane traction
stresses.
It is also found that the substratum stiffness sensed by applying out-of-plane
traction stresses may be up to 10 times larger than the stiffness sensed by
applying in-plane traction stresses.
%


\section{Introduction}

Adherent cells exert mechanical forces on the extracellular matrix (ECM) to
regulate adhesions, propel their migration\cite{li:gu:chi:05} and to sense the
ECM stiffness by a process generally referred as mechanosensing
\cite{engler2006,dis:jan:wan:05}.  In our organism, cells are often embedded in
three-dimensional (3D) ECMs, which they deform in all spatial directions by
generating three-dimensional forces in order to migrate\cite{bloometal:wir:08}.  
Even cells that form stable two-dimensional (2D) monolayers, such as vascular
endothelial cells, are known to exert three-dimensional traction forces
\cite{huretal:09} both in isolation and confluency \cite{huretal:12}.  The
ability of cells to apply three-dimensional forces on two-dimensional layers is
also important for the extravasation of leukocytes during the immune response
\cite{rabodzey:08} and for cancer cell invasion \cite{Poinclouxetal:11,
Aungetal:12}.

The past few years have witnessed the development of several 2D traction force
microscopy (TFM) methods, which allow investigators to measure only the
in-plane (tangential) components of the traction stresses generated by cells
adhering to plane substrata 
\cite{dembo1996,Munevar2001,butler2002,delalamo2007,Sabass:2008,trepat2009physical}.
More recently, 3D TFM methods have been introduced allowing for the
determination of the out-of-plane (normal) component of the traction stresses
\cite{huretal:09,Maskarinecetal:09}.
However, these 3D TFM methods are based on numerical calculations performed on
large volumetric grids, which can limit their ability to deal with large
experimental sets containing many samples.
Furthermore, some of these techniques \cite{Maskarinecetal:09} rely on the
acquisition of thick $z$-stacks containing a large number of confocal images,
which not only subjects the cells to high levels of laser radiation, but also
constrains the temporal resolution of the measurements.

The realization that cells generate out-of-plane traction stresses even when
they adhere to plane surfaces \cite{huretal:09} has prompted an increasing
demand for the characterization of the error that the widely used 2D TFM
methods may have incurred by neglecting these stresses.
This would require to quantify the error of 2D TFM as a function of
experimental parameters such as cell size, cell shape, substratum thickness and
substratum Poisson ratio (see \S \ref{sec:comp2d3d} for details).  Because
existing 3D TFM methods are not well suited for parametric studies, the error
of 2D TFM has not been characterized yet.

The effect of the normal traction stresses on ECM mechanosensing is another
aspect related to 3D traction stresses that remains unexplored.
Since normal stresses deform the substratum axially while tangential ones shear
it, and these two types of deformation are different from each other, it is
reasonable to question if cells can sense different ECM mechanical properties
by imparting tangential or normal stresses.
Thick isotropic gels have their shear elastic modulus proportional to their
axial Young's modulus and, thus, they exhibit similar stiffness under both
tangential and normal traction stresses \cite{landau}.
However, the same cannot be expected to hold for thin substrata, where the
contact with a rigid surface (\eg a glass coverslip {\it in vitro} or bone
tissue {\it in vivo}) breaks the macroscopic isotropy
of the system as a whole, even if the microscopic material properties of
the gel remain isotropic.
The elastic response to tangential traction stresses in thin substrata has been
previously studied using 2D TFM approaches \cite{maloney:08,linetal:10} but we
still lack information about the response to normal traction stresses.
Harnessing the different values of the elastic modulus sensed by the cell by
tangential and normal traction stresses could provide some control over the
cellular response to an ECM with given mechanical properties.

The aim of this paper is to address these open questions and to improve
existing 3D TFM approaches by introducing a new traction cytometry method based
on an exact analytical three-dimensional solution to the elastic equilibrium
equation for a substratum of finite thickness.
In this method, it is sufficient to measure the 3D deformation of the
substratum on a thin layer near its surface.  Therefore, the cells are exposed
to lower levels of laser radiation and a higher temporal resolution is
obtained.
The analytical solution derived in this study is a general explicit formula
that can be recycled for different sets of deformation measurements, resulting
in an improved computational efficiency compared to existing 3D TFM methods.
The new methodology is described in section \ref{sec:methods} and its
performance is illustrated in section \ref{sec:defostress} for chemotaxing {\it
Dictyostelium} cells.
Section \ref{sec:comp2d3d} compares the results of 2D and 3D TFM methods for a
synthetic 3D deformation field that simulates the deformation patterns measured
in the experiments.  
Specifically, section \ref{sec:err} defines the error of 2D TFM methods and
analyzes its dependence on the experimental parameters in order to obtain the
combinations of these parameters that minimize the error.
A spectral comparison of 2D and 3D TFM methods that does not presume any
synthetic shape for the deformation field is given in section \ref{sec:spec}.
This spectral comparison is extended in section \ref{sec:deep} to study
mechanosensing of substratum stiffness and thickness by 3D traction stresses. 
%


\section{Materials, Methods \& Analysis}
\label{sec:methods}
%

\subsection{\emph{Dictyostelium} Cell Culture}
\label{sec:cells}

\emph{Dictyostelium} cells were grown under axenic conditions in HL5 growth
medium in tissue culture plates {as described in \cite{meilietal:99}.  We
used wild type (\emph{WT}) cells (Ax3).  Aggregation competent cells were
prepared by pulsing a $5\times10^6$  cells/ml suspension in Na/K phosphate
buffer ($9.6$~mM KH$_2$PO$_4$, $2.4$~mM Na$_2$HPO$_4$, pH~$6.3$) with cAMP to a
concentration of $30$~nM every $6$~min for $6$~h.  Cells were seeded onto the
functionalized polyacrylamide substratum and allowed to adhere. A drawn glass
capillary mounted on a micromanipulator served as the source of chemoattractant
($150$~$\mu$ M cAMP in an Eppendorf femtotip, Eppendorf, Germany).  
%

\subsection{Polyacrylamide Gel Preparation}
\label{sec:gel}

$12$--mm diameter and $50$--$100$ $\mu$m thick polyacrylamide gels were
fabricated on $22$--mm square $\#1$ glass coverslips using essentially the
procedure described by Beningo \etal \cite{Beningo2002}.  To improve the signal
to noise ratio of the $z$-stacks, the polyacrylamide gel was fabricated as two
sequential layers with the bottom one containing no beads and
the top one containing $0.03\%$ carboxylate modified yellow
latex beads with $0.1$ $\mu$m diameter (Fluospheres, Invitrogen, Carlsbad CA).  
The two layers were verified to adhere well to each other by performing
supporting experiments using gels where the beads were contained in the bottom
layer.
The square coverslips with the two layered gels were mounted in Petri
dishes with a circular opening in the bottom using silicon grease (Dow Corning,
Midland, Michigan).  We crosslinked collagen I to the surface using
Sulfo-SANPAH (Thermo Sci, Rockford, Il) (1 mM in HEPES buffer (pH 8.5). After
UV activation and washing thoroughly, $0.25$ mg/ml collagen protein in HEPES
were added and the gels were incubated overnight at room temperature.  After
washing the gels were stored with buffer 
(9.6 mM KH2PO4, 2.4 mM Na2HPO4, pH 6.3, same buffer used in the
experiments)
and antibiotic 
(40$\mu$M Ampicillin)
for use within a week.
Substratum thickness was measured by locating the top and bottom planes of the
gel and subtracting their $z$ positions.  The top plane was found by maximizing
the number of in-focus pixels of cell outlines as described by del \'Alamo
\etal \cite{delalamo2007}
The bottom plane was found in a similar manner by focusing on streaky patterns
left on the surface of the glass coverslip during treatment for gel attachment. 
%

\subsection{Microscopy}
\label{sec:microscopy}
%
The 3D deformation of the polyacrylamide substratum was determined from the
displacements of embedded $0.1$--$\mu$m fluorescent beads as described in
section \ref{sec:PIV} (see figure \ref{fig:speckles}).  Time-lapse sequences of
$z$-stacks of fluorescence images were acquired using a 
3I Marianas spinning disk confocal system.  It consist of a Zeiss Observer
Z1 body with a 63X oil immersion lens, a Yokogawa CSU-X1 spinning disk head and
a Roper Quant EM 512 SC camera for image capture. The pinholes have a diameter
of 50 $\mu$m.  The measured full-width half maximum of the point spread
function of the system was measured to be $1.4\, \mu$m.
Each $z$-stack contained $8$--$12$ images separated a vertical distance $\Delta
z = 0.5$ $\mu$m.  The images were collected using an $63X/1.4$ oil lens through
a spinning disk confocal head.  The $xy$ position and the shape of the cells
was recorded with an additional single differential interference contrast (DIC)
transmitted light image for each time point.
%

\subsection{Identification of cell contours}
\label{sec:cellcontour}
%
Cell outlines were determined from the DIC images of the free surface of the
substratum captured as described in \S \ref{sec:microscopy}.  Image processing
was performed with MATLAB (Mathworks Inc, Natick, MA) as described in previous
works by our group \cite{delalamo2007,AlonsoLatorre2010,Bastounis11}.
Static imperfections were removed from individual images using the average of
the image series. A threshold was applied to the resulting images to extract
the most intense features, which were refined using two consecutive image
dilations and erosions with structuring elements of increasing size. The sets
of connected pixels were detected and their holes were filled.
%

\subsection{Measurement of Three-dimensional Substrate Deformation}
\label{sec:PIV}
%
The deformation of the substratum was measured in three dimensions by
cross-correlating each instantaneous fluorescence $z$-stack obtained with the
confocal microscope (see \S \ref{sec:microscopy}) with a reference $z$-stack in
which the substratum is not deformed (see figure \ref{fig:speckles}). 
In each experiment, the undeformed image was obtained by waiting for the cell
to move out of the field of view, which takes approximately 10 minutes as {\it
Dictyostelium} cells are highly motile.
The comparison between the deformed and undeformed (reference) conditions was
performed by dividing each instantaneous and reference $z$-stacks into 3D
interrogation boxes and optimizing the 3D cross-correlation between each pair
of interrogation boxes.  Sub-pixel resolution was attained by 
tri-quadratic
polynomial interpolation of the image correlation function.
The image correlation codes were validated against the 3D single-particle
tracking codes developed by Hur \etal \cite{huretal:09}, showing very good
agreement.
Figure \ref{fig:s2n} in the Supporting Information shows a representative example of the 3D
cross-correlation function between interrogation boxes from the deformed and
undeformed images. 
The signal-to-noise ratio of the measurement is defined as the ratio between
the maximum value of the cross-correlation and the second highest local
maximum.  
The measurement floor estimates the lowest deformation that could be
accurately measured by our method.  It was defined from the standard deviation
of the deformation measured far away from the cell, where this deformation
should be equal to zero.
The horizontal size of the interrogation boxes was chosen to balance
resolution, which decreased with box size, 
measurement floor, which decreased with box size,
and signal-to-noise ratio, which increased with box size (see table
\ref{tab:s2n}).  We chose the smallest box size that provided a signal-to-noise
ratio greater than $2.5$
which resulted in a box with $24 \times 24$ pixels in the $x$ and $y$
directions, and led to a Nyquist spatial resolution of $2.1\; \mu m$. 
Further increases in box size led to higher signal to noise ratios but did
not decrease the measurement floor appreciably.  Typical values of deformations
exerted by the cells ranged between 30 and 50 times the measurement floor (see
figure \ref{fig:examplecell}).
In addition to these considerations, phototoxic effects need to be taken into
account when choosing the vertical size of the interrogation boxes, $\Delta z$,
because the level of laser radiation transmitted to the cells increases with
the number of slices per $z$-stack.  In our experiments, $\Delta z = 8$ was the
minimum box height that allowed for meaningful deformation measurements.  Thus,
we acquired $z$-stacks having between 8 and 12 slices, which enabled us to
record time lapse sequences of stacks with a time resolution of 4 seconds and
durations of up to $30$ minutes with no apparent phototoxic effects.
%

\subsection{Calculation of Three-dimensional Traction Stresses}
\label{sec:calcu}
%

We computed the three-dimensional deformation field in the whole substratum by
solving the elasticity equation of equilibrium for a linear, homogeneous,
isotropic three-dimensional body characterized by its Poisson's ratio,
$\sigma$,
\begin{equation}
%
(1 - 2\sigma) \nabla^2 \mathbf{u} + \nabla (\nabla \cdot \mathbf{u})= 0,
\label{eq:elastostatic}
%
\end{equation}
where the inertial stresses have been neglected.  This assumption is justified
by estimating that the ratio between the inertial and elastic stresses is of
order $ \rho l^2 / E T^2 \sim 10^{-12}$, where $\rho \sim 10^3 kg m^{-3}$ and
$E \sim 10^3 Pa$ are respectively the density and Young modulus of the
substratum, $l \sim 2\times 10^{-5}  m$ is the length of the cell and $T \sim
20 s$ is the characteristic timescale of pseudopod protrusion and retraction
\cite{meili2010}.  We sought for solutions of this equation that are periodic
in both horizontal directions and expressed them using a discrete Fourier
series,
\begin{equation}
%
\mathbf{u}(x,y,z) = \sum_{m=-M/2}^{M/2-1}\sum_{n=-N/2}^{N/2-1} \mathbf{\widehat
u}_{mn}(z) \exp(i \alpha_m x) \exp(i \beta_n y),
%
\end{equation}
where $\alpha_m = 2 \pi m / L$, $\beta_n = 2 \pi n/ W$ are respectively the
wavenumbers in the $x$ and $y$ directions, while $L$ and $W$ are respectively
and the spatial periods of the domain.  The numbers of Fourier coefficients in
each direction are determined from the grid of interrogation
boxes using the Nyquist criterion, yielding $M = L/\Delta x$ and $N = W /
\Delta y$.  Figure \ref{fig:sketch} outlines the problem configuration.
For the sake of brevity and without loss of generality, we will drop the
subindices $m$ and $n$ of $\alpha$ and $\beta$ in what follows (\ie $\alpha =
\alpha_m$ and $\beta = \beta_n$).

The boundary conditions to eq. \ref{eq:elastostatic} are partially set up by
imposing zero displacements at the base of the substratum, i.e.
$\mathbf{u}(x,y,0)=0$ (see fig. \ref{fig:sketch}), since the glass coverslip is
much stiffer than the
gel, and the gel is firmly adhered to the coverslip.  A general solution
to eq.  \ref{eq:elastostatic} that is compatible with zero displacements at the
bottom of the gel was determined by del \'Alamo \etal \cite{delalamo2007}.  
More recently, an elegantly simple solution to the 2D problem that only
requires computing two scalar functions was found by Trepat \etal
\cite{trepat2009physical}.  However, here we follow the more general
formulation by del \'Alamo \etal \cite{delalamo2007}, which we can easily
extend to three dimensions and which is also relatively simple.
We express the Fourier transform of the deformation vector as 
\begin{equation}
%
\mathbf{\widehat u}_{mn}(z) = 
{\left[\begin{array}{c|c|c} 
{\bf U}_{mn}(z)  & {\bf V}_{mn}(z) & {\bf W}_{mn}(z) 
\end{array}\right]}
\cdot
d_z{\bf \widehat u}_{mn}^0 = \mathcal{U}_{mn}(z) \cdot d_z{\bf \widehat
u}_{mn}^0,
\label{eq:gensol}
%
\end{equation}
where ${\bf U}_{mn}$, ${\bf V}_{mn}$, ${\bf W}_{mn}$ are three fundamental
solutions of the problem (given in the Supporting Information).  Thus, $\mathcal{U}_{mn}(z)$
is the resolvant matrix.  The vector $d_z{\bf \widehat u}_{mn}^0$ contains the
$z$-derivatives of the displacements' Fourier coefficients at $z=0$, which are
unknown {\it a priori}.  The same solution was found independently by three
members of our team using Maple (Maplesoft, Waterloo, ON, Canada).

\subsubsection{Boundary Conditions on the displacements at $z=h$}

The three unknown elements of $d_z{\bf \widehat u}_{mn}^0$ can be determined by
imposing the three-dimensional displacements measured at $z=h$ as boundary
conditions (see fig. \ref{fig:sketch}).  For that purpose we determine the
Fourier coefficients of the measured displacements, $\mathbf{\widehat
u}_{mn}(h)$, and invert equation \ref{eq:gensol} particularized at $z=h$ to
arrive at
%
\begin{equation}
%
\mathbf{\widehat u}_{mn}(z) =  \mathcal{U}_{mn}(z) \cdot
\mathcal{U}^{-1}_{mn}(h) \cdot \mathbf{\widehat u}_{mn}(h).
\label{eq:uFou}
%
\end{equation}
Explicit analytical expressions for the nine elements in $\mathcal{U}_{mn}(h)
^{-1}$ are given in the Supporting Information.  
As noted by Butler \etal \cite{butler2002}, working in the Fourier domain allows
us to perform this inversion exactly and without regularization, which can be a
delicate issue in methods that work in the physical domain \cite{dembo1996}.
Analytical differentiation of the resolvant matrix yields explicit formulae for
the $z$-derivatives of the displacements,
\begin{equation}
%
d_z \mathbf{\widehat u}_{mn}(z) =  d_z \mathcal{U}_{mn}(z) \cdot
\mathcal{U}^{-1}_{mn}(h) \cdot \mathbf{\widehat u}_{mn}(h).
\label{eq:dzuFou}
%
\end{equation}

\subsubsection{Determination of the normal and tangential traction stresses on
the deformed surface of the substratum}
\label{sec:tannor}

%
This section employs the solution to the elastostatic equation
(\ref{eq:elastostatic}) to determine the Green's function $
\widehat{\mathcal{G}}_{mn}$ that relates the 3D traction stress vector exerted
by the cell on the surface of the gel, $\boldsymbol{\tau}$, to the measured
displacements, $\mathbf {u}$.  In Fourier space, this relation is given by 
%
\begin{equation}
%
\widehat {\boldsymbol \tau}_{mn}(h) = \widehat{\mathcal{G}}_{mn} \cdot
\widehat{\mathbf{u}}_{mn}(h).
\label{eq:Green}
%
\end{equation}
%
The Green's function is obtained in two steps:

\paragraph{Determination of the vector normal to the deformed surface.}
%
The stresses on the surface of the substratum are given by by $\boldsymbol \tau
= \mathcal{T} \cdot \mathbf{n}$, where $\mathbf{n}$ is the unit vector normal
to the surface of the substratum, and $\mathcal{T}$ is the three-dimensional
stress tensor.
Assuming that the deformed surface of the gel is given by $z = h + w(x,y,h)$,
the vector normal to the deformed surface is
$$
\mathbf{n} \approx \frac {\left( \partial_x w, \partial_y w, 1 \right)} 
{\sqrt{ \left(\partial_x w\right)^2 + \left(\partial_y w\right)^2 + 1
}}.
$$
The normal stresses on the deformed substratum are then given respectively by
$$
\boldsymbol {\tau}_n = \left(\mathbf{n}^T \cdot \mathcal{T}\cdot
\mathbf{n}\right) \mathbf{n}
\; \; \; \; \; \;
\mbox{and}
\; \; \; \; \; \;
\boldsymbol {\tau}_t = \mathcal{T}\cdot \mathbf{n} - \boldsymbol {\tau}_n. 
$$
Consistent with the small-deformation approximation \cite{landau}, the values
of $\partial_x w$ and $\partial_y w$ in our experiments are typically small, in
the range $0.05-0.1$.  Thus, $\mathbf{n} \approx (0,0,1)$, and the normal and
tangential stresses can be approximated by $\boldsymbol \tau_n =
(0,0,\tau_{zz}) $ and $\boldsymbol \tau_t = (\tau_{xz},\tau_{yz},0)$, similar
to the undeformed conditions.  Likewise, the normal and tangential deformation
can be expressed as $\mathbf u_n = (0,0,w)$ and $\mathbf u_t = (u,v,0)$.
%

\paragraph{Application of Hooke's Law.}
%
The stress vector can be obtained from the calculated displacements and their
$z$-derivatives by applying Hooke's law in Fourier space,
%
\begin{equation}
%
\widehat {\boldsymbol \tau}_{mn}= 
\left[\begin{array}{c}
\widehat{\tau_{xz}} \\ \widehat{\tau_{yz}} \\ \widehat{\tau_{zz}}
\end{array}\right]_{mn} =
\mathcal{H}_{mn}(E,\sigma) \cdot
\left[\begin{array}{c}
\widehat{\mathbf{u}}_{mn}(h) \\
\hline
d_z \widehat{\mathbf{u}}_{mn}(h)
\end{array}\right], 
%
\end{equation}
where the $6\times3$ matrix $\mathcal{H}_{mn}$, which only depends on the
material properties of the substratum ($E$ and $\sigma$) and the wavenumbers of
each Fourier mode ($\alpha_m$, $\beta_n$), is given in the Supporting Information.  Plugging
the solution for $d_z \widehat{\mathbf{u}}_{mn}(h)$ obtained by applying the
boundary conditions (\ie equation \ref{eq:dzuFou} at $z=h$ ), we obtain the
Green's function,
%
\begin{equation}
%
\widehat{\mathcal{G}}_{mn}=
\mathcal{H}_{mn} \cdot
\left[\begin{array}{c}
I \\
\hline
d_z \mathcal{U}_{mn}(h) \cdot \mathcal{U}_{mn}(h) ^{-1}
\end{array}\right]. 
\end{equation}

\paragraph{Determination of the Strain Energy Exerted by the Cells on the
Substrate.}
%
The strain energy deposited by the cell on the substratum is equal to the
mechanical work exerted by the cell and, thus, it can be easily determined
using the principle of virtual work \cite{butler2002,landau}.  
In the 3D case, it is convenient to decompose the strain energy in its normal
($U_n$) and tangential ($U_t$) components,
%
\begin{align}
U_t = \frac 1 2 \int_0^L \int_0^W \boldsymbol{\tau}_t(x,y,h) \cdot \mathbf
u_t(x,y,h) \dd x \dd y
&
\approx \frac 1 2 \int_0^L \int_0^W [\tau_{xz} u + \tau_{yz} v ](x,y,h)
\dd x \dd y,
\\
U_n = \frac 1 2 \int_0^L \int_0^W \boldsymbol{\tau}_n(x,y,h) \cdot \mathbf
u_n(x,y,h) \dd x \dd y
&
\approx \frac 1 2 \int_0^L \int_0^W [\tau_{zz} w] (x,y,h) \dd x \dd y,
\end{align}
%
so that $U_s =  U_t + U_n$ is the total strain energy.  
These two components can be interpreted as the mechanical work exerted by the tangential and normal traction stresses.
\section{Results}

\subsection{Application of 3D Fourier Traction Force Microscopy to Migrating Amoeboid Cells}
\label{sec:defostress}
%

Consistent with previous 2D studies
\cite{delalamo2007,AlonsoLatorre2010,Bastounis11}, we found that cells contract
the substratum from front and back towards the center of the cell.  However,
the 3D measurements revealed that the deformation and stresses in the direction
perpendicular to the substratum cannot be neglected, as they can be as large as
the tangential ones.  
These observations are illustrated in figure \ref{fig:examplecell}, which
displays an example of the deformation and stress fields caused by a migrating
{\it Dictyostelium} cell on the surface of the substratum.  The cell in this
figure is the same one represented in figure \ref{fig:speckles} and is
migrating from left to right.  
As a matter of fact, for the specific cell and instant of time depicted in
figure \ref{fig:examplecell}, the peak normal deformation and stresses are even
higher than the peak tangential ones.  The corresponding normal strain energy,
$U_n = 4.7\, nN \mu m$, is also greater than the tangential one, $U_t = 1.6\,
nN \mu m$ (note that $1\, nN \mu m = 1\, fJ$).  

The 3D measurements also indicate that, during migration, the cells not only
tugged the substratum horizontally but also pulled upwards (away from the
substratum) along their front and back while pushing downwards right under
their center (see fig. \ref{fig:examplecell}\ccc, \fff).  These results are in
good agreement with previous 3D traction stress measurements \cite{sung2009}.

\subsection{Comparative analysis of 3D and 2D traction force microscopy methods}
\label{sec:comp2d3d}

This section compares the present 3D TFM method with previous 2D TFM methods
that assumed the substratum to be infinitely thick and neglected the vertical
displacements \cite{dembo1996,butler2002}, and with 2D TFM methods that
accounted for the finite thickness of the substratum but neglected either the
vertical displacements \cite{trepat2009physical} or the vertical traction
stresses at $z=h$ \cite{delalamo2007}.  
The comparison is simplified because, under the small deformation-approximation
(see \ref{sec:tannor}), the vector normal to the free surface of the substratum
is the same in two and three dimensions.  Thus, the $x$--$y$ components of the
2D traction stress vector,
$
\boldsymbol \tau^{2D} = (\tau_{xz}^{2D}, \tau_{yz}^{2D}, 0)
$,
are homologous to those obtained by 3D TFM, where 
$
\boldsymbol \tau^{3D} = (\tau_{xz}^{3D}, \tau_{yz}^{3D}, \tau_{zz}^{3D})
$.
We will thus reduce our analysis to a direct comparison between
$\tau_{iz}^{2D}$ and $\tau_{iz}^{3D}$ for $i = x,y,z$.  Note that in general
$\tau_{iz}^{2D} \ne \tau_{iz}^{3D}$ because of the different boundary
conditions employed in each case to solve equation \ref{eq:elastostatic}.

\subsubsection{Synthetic-field analysis of 3D and 2D traction force microscopy
methods}
\label{sec:err}

In order to systematically study the errors in 2D TFM, we calculate the
traction stresses generated by a synthetic deformation field applied on the
surface of the substratum,
%
\begin{align}
%
u(x,y,h) = & - \frac {U_0} {\sqrt{(x-\Delta)^2 + y^2}} + \frac {U_0}
{\sqrt{(x+\Delta)^2 + y^2}},
\label{eq:usyn}
\\
v(x,y,h) = & 0, \\
w(x,y,h) = & \frac {W_0}/2 {\sqrt{(x-\Delta)^2 + y^2}} + \frac {{W_0}/2}
{\sqrt{(x+\Delta)^2 + y^2}} - \frac {{W_0}} {\sqrt{x^2 + y^2}},
\label{eq:wsyn}
%
\end{align}
which is plotted in figures \ref{fig:synthe}(\aaa)-(\ccc).  
This synthetic field was chosen because it resembles the deformation pattern
caused by migrating amoeboid cells in our experiments (see figure
\ref{fig:examplecell}).  It consists of three deformation poles whose positions
are aligned in the direction parallel to the cell's longitudinal axis (\ie the
$x$-axis).  The front and back poles are placed at a distance $2 \Delta$ to
each other.  They contract the substratum tangentially towards their midpoint,
and pull away from substratum in the direction normal to the free surface.  The
central pole presses down the substratum in the normal direction.  The average
deformation created by these three poles is zero in all directions so that
static equilibrium is fulfilled.

In order to characterize the dependence of the error on the ratio of
tangential to normal deformation, we let $u$ and $w$ be proportional to two
different lenghtscales, $U_0$ and $W_0$. Once we set the value of the ratio
$W_0 / U_0$,%
the actual magnitude of $U_0$ is irrelevant for the purpose of comparing 2D and
3D TFM equations because the elastostatic equation and the constitutive
equations of the substratum are linear.  The same applies to the magnitude of
the substratum's Young modulus. Therefore, we set $U_0 = 1$ (units length) and
$E = 1$ (units force per unit area) for simplicity. 
The deformation field caused by each pole was chosen to decay as the inverse
distance to each pole because this rate of decay is known to result from
applying a point force (\ie a Dirac delta traction stress) at the pole when $h
\gg \Delta$ \cite{landau}.  
This particular choice allowed us to verify the correctness of our calculations
and is immaterial to the comparison of 2D and 3D methods presented below.
Figure \ref{fig:synthe03gau} in the Supporting Information shows that the same
comparative results are obtained for other choices of the deformation field.
The only relevant non-dimensional parameters for this comparison are the ratio
$\Delta/h$, which can be interpreted as a surrogate for cell length relative to
substratum thickness, the Poisson's ratio of the gel,%
and the ratio of normal to tangential deformation, $W_0/U_0$.  Note that
$W_0/U_0$ is somewhat related to the vertical aspect ratio of the cell; flat
cells are geometrically constrained to generate low values of $W_0/U_0$ while
round cells may generate high values of $W_0/U_0$.
In order to quantify more precisely the accuracy of 2D TFM, and its dependence
on the Poisson's ratio, $W_0/U_0$ and $\Delta / h$, we define the error in the tangential
stresses as $\mathcal{E}_t = (\mathcal{E}_{t,x} +
\mathcal{E}_{t,y})/2$, where
%
\begin{equation}
%
%
\mathcal{E}_{t,x}(\sigma,W_0/U_0,\Delta/h) = \left\{ \frac{\int \int
\left[\tau^{2D}_{xz}(x,y,z=h)
- \tau^{3D}_{xz}(x,y,z=h) \right]^2\dd x \dd y}
{\int \int \tau^{3D}_{xz}(x,y,z=h)^2  \dd x \dd y}
\right\}^{1/2},
\label{eq:relerrt}
%
\end{equation}
and $\mathcal{E}_{t,y}$ is defined in a similar way.
The relative error in the normal stresses is defined as
\begin{equation}
%
%
\mathcal{E}_{n}(\sigma,W_0/U_0,\Delta/h) = \left\{ \frac{\int \int
\left[\tau^{2D}_{zz}(x,y,z=h)
- \tau^{3D}_{zz}(x,y,z=h) \right]^2\dd x \dd y}
{\int \int \tau^{3D}_{zz}(x,y,z=h)^2 \dd x \dd y}
\right\}^{1/2}.
\label{eq:relerrn}
%
\end{equation}
%
These errors are normalized so that they are equal to one when $\tau^{2D} = 0$
and $\tau^{3D} \ne 0$.
%

\paragraph{
Influence of the Poisson ratio and cell length on the error of 2D TFM.
}
%
Based on the evidence that the level of normal deformation is comparable to
the level of tangential deformation (see $\S$ \ref{sec:defostress}), our
analysis will focus first on the dependence of the error on $\sigma$ and
$\Delta /h$ for the case $W_0/U_0 = 1$.
For each pair of values of $\Delta / h$ and Poisson's ratio, one can calculate
a traction stress map from the boundary conditions specified by the synthetic
deformation field in eqs. \ref{eq:usyn}-\ref{eq:wsyn} (see figure
\ref{fig:sketch}).   
Figure \ref{fig:synthe}(\ddd)-(\fff) shows the traction stresses obtained by
3-D Fourier TFM for $\sigma = 0.3$ and $\Delta / h = 1/2$, which are
representative values of the experimental conditions in TFM experiments
\cite{delalamo2007,meili2010,Bastounis11} 
The lower panels in that figure display the stresses obtained from the
synthetic deformations by 2D TFM methods on finite-thickness substrata
\cite{delalamo2007,trepat2009physical}.
These stresses are obtained by replacing the boundary condition (\ref{eq:wsyn})
with either $w(x,y,h)=0$ (panels \ggg--\iii, ref. \cite{trepat2009physical}),
or with $\tau_{zz}(x,y,h)=0$ (panels \jjj--\lll, ref. \cite{delalamo2007}).  

A visual inspection of figure \ref{fig:synthe} provides overall qualitative
information about the accuracy of 2D TFM in comparison to the 3D
approach introduced in this work.  Both 2D methods produce similar results and
are able to reproduce the tangential contractile stresses along the $x$
direction ($\tau_{xz}$) at the front and back poles (fig.  \ref{fig:synthe}\ddd,
\ggg, \jjj).  However, the 2D methods miss the tangential expansive stress
caused by the central pole's pushing down into the substratum, which is due to
the Poisson's effect.
The accuracy of the 2D methods is somewhat worse for the tangential stresses in
the $y$ direction ($\tau_{zy}$, figure \ref{fig:synthe}\eee,\,\hhh,\,\kkk)
because these stresses appear exclusively due to the Poisson's effect caused by
the deformations prescribed in the other two directions.
Finally, and as expected, 2D TFM severely underpredicts the
normal traction stresses ($\tau_{zz}$, figure
\ref{fig:synthe}\fff,\,\iii,\,\lll). 
The 2D methods that impose $w=0$ on the surface of the substratum
\cite{trepat2009physical} only capture the normal stresses caused by the
tangential deformation due to the Poisson's effect.  And obviously, the
$\tau_{zz} = 0$ condition \cite{delalamo2007} leads to zero normal stresses on
the surface of the substratum.

Figure \ref{fig:errsig1D}(\aaa) displays the tangential errors as a function of
the Poisson's ratio for two extreme values in which $\Delta/h \ll 1$ ($0.01$)
and $\Delta/h \gg 1$ ($100$).  
It is worth studying these limits in detail because they contain zones of low
$\mathcal{E}_t$, which can be used to guide the design of 2D TFM experiments
that are accurate independent of the normal stresses exerted by the cell.
The typical range of Poisson's ratios of the gels employed in traction force
microscopy, $0.3 \le \sigma < 0.49$
\cite{chippada2010simultaneous,takigawa1996poisson,li1993new}, is indicated by
the shaded regions in the plot.  

For low values of $\Delta /h $ (blue lines in fig. \ref{fig:errsig1D}\aaa), the
length of the cell is much smaller than the thickness of the substratum.  In
this case, Boussinesq's elastostatic solution for infinitely thick substrata
\cite{butler2002} becomes equal to the two finite-thickness solutions
\cite{delalamo2007,trepat2009physical}, which are also approximately equal to
each other consistent with figure \ref{fig:synthe}.
Hence, the tangential errors of all three 2D TFM methods are the same for
$\Delta/h \ll 1$.
In the range of interest of Poisson's ratios, $\mathcal{E}_t$ from 2D TFM
methods is up to 
$\approx 35\%$ but this error
decreases sharply as the Poisson's ratio approaches the incompressible limit,
$\sigma = 0.5$. 
This behavior is explained by recalling that the tangential stresss/strain
fields in the Boussinesq solution decouple from the normal ones in that limit
(ref.  \cite{landau}, page 25, eq.  8.19).  Ideally, it would not be necessary
to measure the normal displacements in order to accurately determine the
tangential stresses under these conditions ( $\sigma = 0.5$, $\Delta \ll h$).
In practice, however, the sharp decrease of $\mathcal{E}_t$ means that this
error remains relatively high for values of the Poisson's ratio close to
$\sigma = 0.5$.  For instance, we obtain that 
$\mathcal{E}_t = 13\%$ for $\sigma = 0.45$.

For high values of $\Delta /h$ (red lines in fig. \ref{fig:errsig1D}\aaa), the
length of the cell is much larger than the thickness of the substratum.  
In consequence, Boussinesq's solution yields errors  $\ge
50\%$ for all values of the Poisson's ratio (fig. \ref{fig:errsig1D}).  
The finite-thickness 2D TFM methods \cite{delalamo2007,trepat2009physical}
yield relatively low tangential errors as long as the substratum is far from
incompressible but their $\mathcal{E}_t$ increases abruptly as the Poisson's
ratio approaches $\sigma = 0.5$.
For reference, this error is 
$\mathcal{E}_t \approx 2 \%$ for $\sigma = 0.3 $ but it becomes
$\mathcal{E}_t \approx 35 \%$ for
$\sigma = 0.45$.
The reason for this behavior can be understood by analyzing the elastostatic
equations (\ref{eq:elastostatic}) for very thin substrata.  In this limit, the
$z$-derivatives are much larger than the $x$,$y$-derivatives and the equations
are simplified to
\begin{align}
(1-2\sigma) \partial_{zz} u + \partial_{xz}w = 0,
\label{eq:thinu}
\\
(1-2\sigma) \partial_{zz} v + \partial_{yz}w = 0,
\label{eq:thinv}
\\
\partial_{zz} w = 0.
\label{eq:thinw}
\end{align}
This simplification shows that, if $\sigma$ is far from $0.5$, the second terms
in equations (\ref{eq:thinu})-(\ref{eq:thinv}) are negligible as they contains
$x$ and $y$ derivatives.  In that case, the normal displacements decouple from
the tangential ones and it is not necessary to measure the former in order to
determine the latter.  
However, as $\sigma$ approaches $0.5$, the factor $(1-2\sigma)$ that multiplies
the first terms in equations (\ref{eq:thinu})-(\ref{eq:thinv}) also becomes
small, the $x$ and $y$ derivatives in those equations cannot be neglected
anymore, and the tangential displacements remain coupled to the normal ones.
Figure \ref{fig:errhh1D} is the dual of figure \ref{fig:errsig1D}.  It
illustrates the dependence of $\mathcal{E}_t$ and $\mathcal{E}_n$ on $\Delta /
h$ for two values of the Poisson's ratio typical of TFM experiments, namely
$\sigma=0.3$ and $\sigma=0.45$
\cite{chippada2010simultaneous,takigawa1996poisson,li1993new}.  
For $\Delta < h$, the tangential error of all 2D TFM methods is independent of
$\Delta / h$ and relatively high ($\mathcal{E}_t \approx 37 \%$ and $13\%$ respectively for $\sigma
=0.3$ and $0.45$, fig. \ref{fig:errhh1D}\aaa). 
For $\Delta>h$, $\mathcal{E}_t$ from Boussinesq's 2D method increases steeply
with $\Delta/h$, reaching values that exceed $100\%$ for $\sigma=0.45$.
As mentioned above, this large error is due again to the assumption of
infinite-thickness substrate made in Boussinesq's solution to the elastostatic
equation.
The $\Delta/h$-dependence of the tangential error of the finite-thickness 2D
TFM methods varies strongly with $\sigma$ for $\Delta > h$, consistent with the
results in figure \ref{fig:errsig1D}(\aaa). 
For $\sigma=0.3$, the tangential error decreases monotonically to zero with
$\Delta / h$, whereas for $\sigma = 0.45$, $\mathcal{E}_t$ reaches a maximum
before decreasing. 

Figure \ref{fig:err2Dsig2D} summarizes the dependence of $\mathcal{E}_t$ on the
Poisson's ratio and $\Delta / h$ for both finite-thickness 2D methods (panel
\aaa) and the Boussinesq method (panel \bbb).
For reference, note that the line plots in figures \ref{fig:errsig1D}(\aaa) and
\ref{fig:errhh1D}(\aaa) are respectively horizontal and vertical 1D cuts of the
contour maps in figure \ref{fig:err2Dsig2D}.
The thick white contours in this figure enclose the regions of the $(\sigma,
\Delta/h)$ domain where $\mathcal{E}_t$ is lower than $10\%$. 
Consistent with figures \ref{fig:errsig1D}(\aaa) and \ref{fig:errhh1D}(\aaa),
we find low errors in two regions:
one corresponding with small cell size compared to substratum thickness
($\Delta \ll h$) and incompressible gel ($\sigma = 0.5$), and another region
corresponding with large cell size compared to substratum thickness ($\Delta
\gg h$) and relatively low Poisson's ratio $(\sigma \lesssim 0.35$).  
The first low-error region ($\Delta \ll h$, $\sigma = 0.5$) is also obtained
for Boussinesq's method but this region turns out to be narrow because
$\mathcal{E}_t$ is locally sensitive to small changes in $\sigma$.  This
sensitivity leads to significant error values within the range of Poisson's
ratios typically encountered in TFM experiments.
The second low-error region ($\Delta \gg h$, $\sigma \lesssim 0.35$) seems to
be more robust for the design of TFM experiments because $\mathcal{E}_t$ shows
a mild local dependence on both the substratum thickness and Poisson's ratio,
particularly around $\sigma = 0.3$.
This region is not observed in Boussinesq's method, in which the assumption of
infinitely thick substrate is incompatible with $\Delta$ being larger than $h$.

For the sake of completeness, the errors of all 2D TFM methods in the normal
direction are shown in figures \ref{fig:errsig1D}(\bbb) and
\ref{fig:errhh1D}(\bbb).  These errors are $\mathcal{E}_t \approx
100\%$ regardless of the boundary condition applied on the gel's surface, and
the Poisson's ratio, reflecting the difficulty of determining the normal
traction stresses without measuring the normal deformation of the substratum.

\paragraph{Influence of the ratio of normal to horizontal deformation on
the error of 2D TFM }
%
Flattened or elongated cells such as epithelial cells or neurons are
geometrically constrained and, thus, they are likely to generate predominantly
tangential deformations along significant portions of their basal surface.  On
the other hand, rounder cells such as cancer cells invading into 3D matrices
exert predominantly normal deformations \cite{Aungetal:12}.
The aim of this section is to characterize the performance of 2D TFM methods
under these different scenarios by quantifying their error as a function of the
ratio of normal to horizontal deformation ($W_0/U_0$ in eqs.
\ref{eq:usyn}-\ref{eq:wsyn}).  As noted above, $W_0/U_0$ is loosely related to
the vertical aspect ratio of the cell, as one can argue that $W_0/U_0 \ll 1$
for flat cells and $W_0/U_0 \gtrsim 1$ for rounder cells. 
Figure \ref{fig:errW}(\aaa) displays the tangential error of 2D TFM with
$w(x,y,h) = 0$ as a function of $\Delta / h$ and $W_0/U_0$.  As expected,
$\mathcal{E}_t$ decreases to zero as $W_0/U_0$ tends to zero and consistent
with figure \ref{fig:err2Dsig2D}(\aaa), this decrease is more gradual in thin
substrata.
Similar to figure \ref{fig:err2Dsig2D}, the isoline $\mathcal{E}_t = 0.1$ has
been outlined with a thick white contour, revealing that $W_0/U_0$ must be
lower than $0.25$ to achieve a tangential error below $10\%$ for $\sigma =
0.45$.
Our analysis suggests that $\mathcal{E}_t$ can be relatively high for the
values $W_0/U_0$ that are often reported in experiments.  For instance, the
normal deformations are approximately twice larger than the tangential ones for
the {\it Dictyostelium} cell reported in figure \ref{fig:examplecell}, whereas
Hur \etal's experiments on endothelial cells are consistent with $W_0/U_0
\approx 1/2$ both for single cells and confluent monolayers
\cite{huretal:09,huretal:12}.
In all these cases the tangential error of 2D TFM exceeds $20\%$ according to
figure \ref{fig:errW}(\aaa). 

Plotting $\mathcal{E}_t$ as a function of $\sigma$ and $W_0/U_0$ for a fixed
value of $\Delta / h$ (figure \ref{fig:errW}\bbb) reveals that, when $W_0/U_0
\ll 1$, the error of 2D TFM is low and relatively independent of the Poisson
ratio. 
However, this error increases and becomes very sensitive to the Poisson ratio
as $W_0/U_0$ is augmented.
Interestingly, we find that the isolines of $\mathcal{E}_t$ in figure
\ref{fig:errW}(\bbb) are well approximated by the formula
%
\begin{equation}
\left. {\frac{W_0}{U_0}}\right|_{\mathcal{E}_t} = \frac {0.13 \mathcal{E}_t} {1
- 2\sigma_{\mathcal{E}_t} }
\label{eq:errform}
\end{equation}
when the error is sufficiently small (\ie $\mathcal{E}_t \le 0.2$).  This
approximation is found to be uniformly valid for $\Delta / h \lesssim 3$, and
provides a simple formula to relate $W_0/U_0$ and $\sigma$ for a given level of
error.  For instance, we establish that $W_0/U_0$ needs to be smaller than
$0.26$ to keep the error below $10\%$ when the Poisson ratio is $\sigma=0.45$.
Alternatively, it can be seen that $\sigma$ needs to be higher than $0.487$ to keep
the error below $10\%$ for $W_0/U_0=1$. 
%

\subsubsection{Spectral analysis of 2D and 3D traction force microscopy methods}
\label{sec:spec}

Fourier analysis of the elastostatic equation (\ref{eq:elastostatic}) provides
a general framework to compare different TFM methods that does not rely on any
assumption for the shape of the deformation field, and which is complementary
to the synthetic-field approach followed in the previous section.
This analysis also allows for a general definition of thin substrata as we
show below.
Figure \ref{fig:kerncomp} displays the Fr\"obenius norm of the Green's
functions used in several TFM methods,
%
\begin{equation}
%
||{\bf \widehat{\mathcal{G}}}(\lambda_x,\lambda_y)|| 
=\sqrt{\mbox{trace}\left[{{\bf
\widehat{\mathcal{G}}}^{H}(\lambda_x,\lambda_y){\bf \cdot \,
\widehat{\mathcal{G}}}(\lambda_x,\lambda_y)}\right]},
\label{eq:norma}
%
\end{equation}
%
where $^{H}$ denotes Hermitian transpose. 
This norm provides a measure of how much a given TFM method amplifies or
reduces a harmonic displacement field consisting of wavelengths $\lambda_x = 2
\pi / \alpha$ and $\lambda_y = 2 \pi / \beta$ in the $x$ and $y$ directions.  
Different to the previous section, where we separately analyze tangential and
normal displacements and stresses, the norm employed here provides an overall
conservative value that combines the tangential and normal components of $\bf
u$ and $\boldsymbol \tau$.
This property is common to most tensorial norms and, therefore, the results
from the analysis are independent of the particular choice of norm for the
Green's function.  In fact, we obtain similar results when using other matrix
norms such as the 1-norm, the 2-norm and the $\infty$-norm.

The wavelengths in figure \ref{fig:kerncomp} are normalized with the thickness
of the substratum, $h$, so that high values of $\lambda / h$ correspond to
spatial features of the displacement field that are long compared to the
substratum thickness, and vice versa.  
Overall, we find that 2D TFM methods underestimate $||{\bf
\widehat{\mathcal{G}}}||$, consistent with the results from the previous
section.  The observed differences are small for low $\lambda / h$ (\ie thick
gels).  
However, the Green's functions of finite-thickness and infinite-thickness TFM
methods diverge significantly for $\lambda / h > 3$, suggesting that the
zero-deformation boundary condition imposed at the base of the substratum is
felt when the lateral extent of the traction stresses becomes larger than
$\approx 3 h$. 
This result provides a natural definition of thin substratum for a given
traction stress field; a substratum can be considered as mechanically thin if
the dominant wavelength in the deformation field measured on its surface,
$\lambda_d$, is longer than $3h$.
Del \'Alamo \etal \cite{delalamo2007} showed that $\lambda_d$
is equal to the cell length in migrating amoeboid cells by
measuring the spectral energy density of the deformations
exerted by the cells.
Although it is possible that $\lambda_d$ varies with cell type and from single
cells to confluent monolayers, the present definition of mechanically thin
substratum can be applied to each condition once the deformation field is
measured.

This divergence appears to increase as the Poisson's ratio approaches the
incompressible limit $\sigma = 0.5$, as confirmed by figure \ref{fig:kerncomp2}
in the Supplementary Information.
Note in particular that the Green's functions from finite-thickness 2D TFM
methods severely underestimate the 3D Green's function for $\lambda / h > 3$.
A possible interpretation for this behavior is that normal stresses applied at
$z=h$ penetrate deeper into the substratum than tangential ones.  This
hypothesis is further analyzed and confirmed in the next section.

\subsection{Normal traction stresses reach deeper into the substratum and
mechano-sense larger elastic moduli than tangential ones}
\label{sec:deep}

Del \'Alamo \etal \cite{delalamo2007} reasoned that the finite-thickness
Green's function diverges from Boussinesq's when the $\mathbf{u = 0}$ boundary
condition imposed at the bottom of the substratum is felt at the substratum's
surface.
Inspection of figure \ref{fig:kerncomp} reveals that divergence occurs for
lower values of $h$ in the case of a normal force than for a tangential force,
suggesting that cells should be able to feel deeper into the substratum by
applying normal forces.

A possible way to quantify this effect is to calculate the apparent elastic
moduli of the substratum to tangential and normal traction stresses, $E_{a,t}$
and $E_{a,n}$ respectively. 
We define these moduli as the norms of the tangential and normal restrictions
of $\bf \widehat{\mathcal{G}}$ for harmonic deformations of wavenumbers
$(\alpha,\beta)$ in a substratum of thickness $h$ normalized with their value
at $h=\infty$.  These restricted operators are given by
$$
{\bf \widehat{\mathcal{G}}_t}(\alpha h,\beta h) =
\left[\begin{array}{ccc} \widehat{\mathcal{G}}_{xu}(\alpha h,\beta h) &
\widehat{\mathcal{G}}_{xv}(\alpha h,\beta h) &
\widehat{\mathcal{G}}_{xw}(\alpha h,\beta h)\\
\widehat{\mathcal{G}}_{yu}(\alpha h,\beta h) &
\widehat{\mathcal{G}}_{yv}(\alpha h,\beta h) &
\widehat{\mathcal{G}}_{yw}(\alpha h,\beta h)\\ 0 & 0 & 0 \end{array}\right], 
$$ and $$
{\bf \widehat{\mathcal{G}}_n}(\alpha h,\beta h)=
\left[\begin{array}{ccc} 0 & 0 & 0 \\ 0 & 0 & 0 \\
\widehat{\mathcal{G}}_{zu}(\alpha h,\beta h) &
\widehat{\mathcal{G}}_{zv}(\alpha h,\beta h) &
\widehat{\mathcal{G}}_{zw}(\alpha h,\beta h) \end{array}\right]
$$
respectively.
For simplicity and without loss of generality, we focus on isotropic
deformation fields with $\alpha=\beta=k/\sqrt{2}$ where $k$ is the modulus of
the wavenumber vector, obtaining
\begin{equation}
E_{a,t}(kh) =  \frac {E ||{\bf
\widehat{\mathcal{G}}_t}(kh/\sqrt{2},kh/\sqrt{2}) ||} {||{\bf
\widehat{\mathcal{G}}_t}(\infty,\infty)|| } 
\; \;\; \;
\mbox{and} 
\; \; \; \;
E_{a,n}(kh) = \frac {E ||{\bf \widehat{\mathcal{G}}_n}(kh/\sqrt{2},kh/\sqrt{2})
||} {||{\bf \widehat{\mathcal{G}}_n}(\infty,\infty)|| }.
\label{eq:apparentE}
\end{equation}
Note that $E_{a,t}$ and $E_{a,n}$ are analogous to the shear and axial elastic
moduli of the substratum when measured by applying traction stresses on its
free surface.
Figures \ref{fig:deep2D}(\aaa)-(\bbb) display two-dimensional contour maps of
$E_{a,t}$ and $E_{a,n}$ as a function of the Poisson's ratio and the substratum
thickness normalized with $k^{-1}$, while figure \ref{fig:deepH}(\aaa) shows
one-dimensional cuts of these maps for two constant representative values of
the Poisson's ratio, $\sigma=0.3$ and $0.45$. 
These data demonstrate that, as argued in the introduction, both $E_{a,t}$ and
$E_{a,n}$ are equal to the bulk Young's modulus of the substratum, $E$, for
large values of $kh$.
However, the apparent elastic moduli increase steeply as $kh$ decreases below
$\approx 2$, indicating that the substratum becomes effectively stiffer as its
thickness decreases.
More importantly, the observed increase is significantly more pronounced in
$E_{a,n}$ than in $E_{a,t}$, which suggests that the response of thin substrata
to normal stresses is stiffer than its response to tangential stresses.
We also find that $E_{a,n}$ increases strongly with the Poisson's ratio,
especially near the incompressible limit, where $E_{a,n}$ can be $>10$ times
higher than $E_{a,t}$.  On the other hand, the apparent shear modulus shows a
mild decrease with $\sigma$.
These different behaviors can be explained by noting that normal deformations
compress the substratum, so it is reasonable to expect that normal forces
disturb the substratum more globally and are more affected by substratum
compressibility than tangential ones. 
An increased apparent shear elastic modulus that qualitatively agrees with our
predictions was previously found by 2D TFM methods \cite{linetal:10}.  However,
the substantially larger increase in axial apparent modulus predicted by our
analysis could not be determined in previous 2D studies.  
Altogether, these results confirm that applying normal traction forces at the
surface of the substratum provides higher sensitivity to the infinitely rigid
bottom than applying tangential traction forces.  Or, in other words, that
cells may feel deeper into the substratum by normal traction forces.  In order
to provide a theoretical measure of the sensing depth by tangential and normal
forces, we use $h_{E_a = 2E}$ which is the substratum thickness at which the
elastic modulus perceived by applying traction forces is twice the Young's
modulus.
Figure \ref{fig:deepH}(\aaa) explains how this sensing depth is calculated and
figure \ref{fig:deepH}(\bbb) displays $h_{E_a = 2E}$ for both tangential and
normal forces as a function of the Poisson's ratio.
Consistent with the results in figure \ref{fig:kerncomp}, we find that the
sensing depth is up to $h_{E_a = 2E} = 2 k^{-1} \approx \lambda/3$, which is
equal to $1/3$ the lateral extent of the traction stresses exerted by the cell.
As expected, normal traction forces lead to larger values of the sensing depth
that increase sharply with the Poisson's ratio.
On the other hand, tangential forcing leads to lower values of the sensing
depth which depend little on the Poisson's ratio.
In particular, within the range of values reported for the Poisson's ratio of
polyacrylamide gels
\cite{chippada2010simultaneous,takigawa1996poisson,li1993new},
$\sigma=0.3-0.49$ (shaded region in figure \ref{fig:deepH}\bbb), our
calculations suggest that normal forces can reach between 2 and 4 times deeper
into the substratum than tangential ones.
%

\section{Discussion and Conclusion}

Adherent cells exert three-dimensional (3D) traction stresses on the
extracellular matrix even when cultured on plane surfaces. 
This study presents a novel traction force microscopy (TFM) algorithm that
provides the 3D traction stresses from measurements of the 3D deformation of a
thin layer below the surface of the substratum.  The new algorithm is based on
an exact analytical solution to the equation of mechanical equilibrium for a
finite-thickness substratum, using the measured displacements as boundary
conditions.
The thus obtained solution provides all the elements of the three-dimensional
strain and stress tensors everywhere in the substratum, not only the traction
stresses in the surface.
We have used this new 3D solution to estimate the error of previous
two-dimensional (2D) TFM methods, which only provide the in-plane (tangential)
components of the traction stress vector and neglect its out-of-plane (normal)
component.
We have also used this solution to characterize the influence of the finite
thickness of the substratum and to evaluate the different apparent elastic
moduli that a cell can sense by applying either tangential or normal traction
stresses.

Hur \etal \cite{huretal:09} developed the first TFM method to measure 3D
traction stresses generated by cells adhering on plane surfaces.
Similar to our study, Hur \etal used the measured 3D deformations at the top of
the substratum as boundary conditions to solve the equation of elastic
equilibrium.  However, instead of finding an exact analytical solution that is
compatible with the measured displacements as is done in our study, they
employed a finite element solver to determine the 3D traction stresses from the
measured deformations.
Subsequently, Hur \etal \cite{huretal:12} extended their method to estimate 3D
cell-cell junction forces in cultured cell monolayers.
Maskarinec \etal \cite{Maskarinecetal:09} proposed an alternate 3D TFM approach
that discretizes the whole substratum into a volumetric mesh on which the 3D
deformations are measured.  All the elements of the 3D strain tensor are then
computed on this mesh from the measured deformations, and Hooke's law is
directly applied to determine all the elements of the stress tensor on each
voxel of the mesh.

The 3D TFM method introduced in this paper has a number of advantages with
respect to previously existing methods.  Similar to Hur \etal
\cite{huretal:09}, the present method only requires acquiring a thin ($\sim 8
\mu m$ thickness) $z$-stack composed of few image planes ($\sim 10$).  
This imaging mode allows for a high temporal resolution as the acquisition time
of a confocal z-stack is roughly proportional to the number of recorded planes.
Imaging in a thin layer also reduces the amount of laser radiation that is
imparted on the cells, thereby minimizing phototoxic effects.
Furthermore, solving the elastostatic equation (\ref{eq:elastostatic}) instead
of directly applying Hooke's law has the advantage of not requiring
$z$-derivatives, which can be relatively inaccurate due to the stretch of the
point-spread function of the optics in the $z$-direction.

An additional advantage of the method presented here is that it employs an
exact analytical solution of the elastostatic equation (\ref{eq:elastostatic}),
which allows for computing the traction stresses in virtually zero time from
the deformation at $z=h$.  This is made possible by working in the Fourier
domain and using Fourier Transform (FFT) routines.
The in-house 3D image correlation techniques employed to determine the
deformations from the microscopy images also employ FFTs, which further
improves the computational efficiency of the present algorithm.
The combined features of our method allow us to completeley calculate the
traction stress field from a raw microscopy image stack in $\approx 30$ seconds
on a laptop computer.  
In comparison, Maskarinec \etal \cite{Maskarinecetal:09} report overall
computational times of $24-48h$ per experiment.  Hur \etal \cite{huretal:09}
report computational times of $5$ mins for their finite element calculation but
this does not include the additional processing time required to measure the
deformation field, which is not reported.

Nevertheless, the most important feature of working with an analytical solution
is not that it saves us computational time.  For the size of the computational
problem involved in a TFM experiment, one could always resort to a faster
computer, or parallelize the software routines and run them in multiple
processors to speed up the computation.
The most valuable aspect of an analytical 3D TFM solution is that it offers us
a comprehensive tool to analyze the influence of all the parameters involved in
the problem.
In this work, we have exploited this important feature to characterize the
error incurred by the 2D TFM methods in calculating the in-plane traction
stresses as a function of cell size, substratum thickness and Poisson's ratio.
The aim of this characterization is to establish the range of experimental
parameter values within which 2D TFM experiments may still be used to
accurately measure the in-plane stresses regardless of the fact that the cell
also generates normal traction stresses.
By analyzing the error of the 2D methods in a synthetic 3D deformation field
that mimics the deformation caused by migrating cells, we have shown that 2D
TFM methods that consider the finite thickness of the substratum
\cite{delalamo2007,trepat2009physical} may perform relatively well when the
substratum is exactly incompressible ($\sigma = 0.5$) and its thickness is
larger than the cell's length. 
However, the error of 2D TFM is found to be very sensitive to small variations
of the Poisson's ratio near the incompressible limit, and to increase
considerably when the Poisson's ratio is slightly smaller than $0.5$.  For
instance, we estimate that a typical experiment performed on a $20-\mu m$-long
cell seeded on a $100-\mu m$-thick substrate of $\sigma=0.45$ yields errors
$\approx 15\%$ in the tangential traction stresses when the vertical
deformations are neglected, and this error is even larger if $\sigma$ is
further decreased. 
We also find a second parametric region resulting in low 2D TFM errors when the
cell length is much larger than the thickness of the substratum and the
Poisson's ratio is lower than $\sigma \approx 0.35$.  In this second region,
the error in the tangential traction stresses shows a mild dependence on both
the substratum thickness and Poisson's ratio, especially near $\sigma = 0.3$,
thereby providing a more robust set of experimental conditions than the first
region discussed above.
It is important to note, nonetheless, that 2D TFM methods that assume an
infinitely thick substratum yield errors $\approx 50\%$ when the cell length
is much larger than the substratum thickness.

Given that the traction stresses are proportional to the Young's modulus of the
substratum, significant efforts are generally devoted to the characterization
of this parameter in TFM experiments.  On the other hand, the Poisson's ratio
usually receives less attention under the general assumption that this
parameter barely influences the traction stresses.
The results of this study call for a reexamination of that assumption, as
they reveal a strong $\sigma$-dependence of both the normal traction stresses
and the error of the tangential traction stresses.
The observed sensitivity to the Poisson's ratio is especially significant for
the nearly incompressible gels that are typically used in TFM experiments
\cite{chippada2010simultaneous,takigawa1996poisson,li1993new}.
For the above example TFM experiment (a $20-\mu m$-long cell seeded on a
$100-\mu m$-thick substrate of $\sigma=0.45$), we estimate that a $10\%$
variation in $E$ leads to a $10\%$ variation in the traction stresses, whereas a
similar error in $\sigma$ (from $0.45$ to $0.495$) leads to a $\approx 38\%$
variation in the traction stresses.
It should be noted that, although this study focused on cells adhered onto 2D
surfaces, the 3D Green's functions derived here could be applied to determine
the traction stresses created by living cells in 3D matrices.  In the past,
this problem has been tackled using finite element methods
\cite{Legantetal:2010}.  We speculate that working with exact analytical
solutions in Fourier space could be advantegeous in terms of convergence and
numerical efficiency, as this formulation renders the matrices involved in the
problem block-diagonal \cite{butler2002}.

Adherent cells have been reported to sense the mechanical properties (\ie
Young's modulus) of the extracellular matrix by applying traction stresses to
it \cite{engler2006,dis:jan:wan:05}.
It is recognized that this sensing process can be affected by the contact of
rigid elements with the extracellular matrix (\eg glass at the bottom of
substratum in an {\it in vitro} culture or bone {\it in vivo},
\cite{senetal:09}), leading the cell to sense apparent stiffness values that
exceed the bulk properties of the ECM.
We defined the apparent elastic modulus of a finite-thickness gel by
comparing the stresses generated by applying displacements at its surface with
those generated in a gel of infinite thickness.  We examined the response of
the gel separately for normal and tangential surface displacements, which
allowed us to obtain both normal and tangential apparent moduli.  Previous 2D
studies had only studied the tangential response \cite{maloney:08,linetal:10}.
Our analysis indicates that the apparent Young modulus of the substratum is
affected by the glass coverslip when the dominant wavelength in the deformation
field, $\lambda_d$, is longer than $3h$.
This result is in general agreement with the 2D results of Lin \etal
\cite{linetal:10}.  However, our three-dimensional analysis indicates that
cells can feel between $2$ and $4$ times deeper into the substratum by
generating normal traction forces compared to when they only generate
tangential stresses.  
We have argued that this difference is due to the axial deformation and
stress generated by normal traction stresses, which penetrate deeper into the
substratum than the shear deformation generated by tangential traction
stresses.
A clear example of this effect is shown in figure \ref{fig:penetra}.
Our analysis also reveals that the apparent stiffness to
normal deformations is considerably higher than
to tangential ones.
In particular, in the range of Poisson's ratios of the gels that are
representative of the extracellular matrix or that are employed to manufacture
substrata for {\it in vitro} TFM experiments
\cite{chippada2010simultaneous,takigawa1996poisson,li1993new}, the normal
apparent elastic modulus can be $>10$ times higher than the tangential one.
The predicted large disparity is interesting because it could allow
investigators to modulate the cell's response to the mechanical properties of
the extracellular matrix by altering the ratio between tangential and normal
traction stresses.  The latter could in turn be controlled via the cell's
aspect ratio by micropatterning the substratum \cite{vincentetal:12}.
Existing theoretical and experimental studies have suggested that cells
mechanosense on length scales similar to that of focal adhesions instead of on
cell length \cite{maloney:08,senetal:09}.
It may be possible to reconcile this evidence with our theoretical study and
the experimental data from Lin \etal \cite{linetal:10} by considering that
although $\lambda_d$ was found to be equal to cell length for {\it Dictyostelium}
cells \cite{delalamo2007}, this dominant wavelength can in principle vary with
cell type and other parameters such as cell density (e.g.  confluent versus non
confluent).  Furthermore, our model tacitly assumes that cells mechanosense by
sampling substratum deformation but recent data suggests that cells may sample
stresses, especially on stiffer substrata \cite{Yip201319}.  Note that, because
stresses come from spatial derivatives of deformations, their dominant
wavelength and thus their associated sensing lengthscale is smaller than
$\lambda_d$.
A potential limitation of the present approach is that, similar to most
existing TFM methods, we assume homogeneous, isotropic, linearly elastic
mechanical properties for the substratum.
While these assumptions facilitate the calculation of an analytical solution to
the elastostatic equation, they are not representative of a number of
conditions that warrant experimental access to cellular traction stresses, such
as in durotactic cell migration on substrata with prescribed stiffness
gradients \cite{Loetal2000}.
In those situations, the present Fourier approach can be generalized using
regular perturbation expansions \cite{kevcole} to obtain new analytical
solutions for the traction stresses that are valid for substrata with slowly
varying properties \cite{AlonsoLatorre2010}.
In conclusion, the 3D TFM method introduced in this paper enables the
theoretical and experimental study of new aspects of cell physiology in more
realistic microenvironments.

\section*{Acknowledgments}

We are indebted to Mr. Jui-Hsien Wang for carefully revising preliminary
versions of this manuscript, and to Drs. Shu Chien, Adam Engler, Sung-Sik Hur
and Shyni Varghese for fruitful discussions concerning 3D TFM.
%


\begin{thebibliography}{10}
\providecommand{\url}[1]{\texttt{#1}}
\providecommand{\urlprefix}{URL }
\expandafter\ifx\csname urlstyle\endcsname\relax
  \providecommand{\doi}[1]{doi:\discretionary{}{}{}#1}\else
  \providecommand{\doi}{doi:\discretionary{}{}{}\begingroup
  \urlstyle{rm}\Url}\fi
\providecommand{\bibAnnoteFile}[1]{%
  \IfFileExists{#1}{\begin{quotation}\noindent\textsc{Key:} #1\\
  \textsc{Annotation:}\ \input{#1}\end{quotation}}{}}
\providecommand{\bibAnnote}[2]{%
  \begin{quotation}\noindent\textsc{Key:} #1\\
  \textsc{Annotation:}\ #2\end{quotation}}
\providecommand{\eprint}[2][]{\url{#2}}

\bibitem{li:gu:chi:05}
Li S, Guan J, Chien S ({2005}) {Biochemistry and biomechanics of cell
  motility}.
\newblock {ANNUAL REVIEW OF BIOMEDICAL ENGINEERING} {7}: {105-150}.
\input{li:gu:chi:05}

\bibitem{engler2006}
Engler AJ, Sen S, Sweeney HL, Discher DE (2006) Matrix elasticity directs stem
  cell lineage specification.
\newblock Cell 126: 677-689.
\input{engler2006}

\bibitem{dis:jan:wan:05}
Discher D, Janmey P, Wang Y (2005) {Tissue cells feel and respond to the
  stiffness of their substrate}.
\newblock Science 310: 1139.
\input{dis:jan:wan:05}

\bibitem{bloometal:wir:08}
Bloom R, George J, Celedon A, Sun S, Wirtz D (2008) {Mapping local matrix
  remodeling induced by a migrating tumor cell using three-dimensional
  multiple-particle tracking}.
\newblock Biophysical journal 95: 4077--4088.
\input{bloometal:wir:08}

\bibitem{huretal:09}
Hur S, Zhao Y, Li Y, Botvinick E, Chien S (2009) {Live Cells Exert
  3-Dimensional Traction Forces on Their Substrata}.
\newblock Cellular and Molecular Bioengineering 2: 425--436.
\input{huretal:09}

\bibitem{huretal:12}
Hur SS, del Alamo JC, Park JS, Li YS, Nguyen HA, et~al. (2012) Roles of cell
  confluency and fluid shear in 3-dimensional intracellular forces in
  endothelial cells.
\newblock Proceedings of the National Academy of Sciences .
\input{huretal:12}

\bibitem{rabodzey:08}
Rabodzey A, Alcaide P, Luscinskas F, Ladoux B (2008) {Mechanical forces induced
  by the transendothelial migration of human neutrophils}.
\newblock Biophysical Journal 95: 1428--1438.
\input{rabodzey:08}

\bibitem{Poinclouxetal:11}
Poincloux R, Collin O, Lizárraga F, Romao M, Debray M, et~al. (2011)
  Contractility of the cell rear drives invasion of breast tumor cells in 3d
  matrigel.
\newblock Proceedings of the National Academy of Sciences 108: 1943-1948.
\input{Poinclouxetal:11}

\bibitem{Aungetal:12}
Aung A, Seo YN, del {\'A}lamo JC, Varghese S (2012) Cancer migration in 3-{D}
  environment.
\newblock Biophysical Journal 102: 706a.
\input{Aungetal:12}

\bibitem{dembo1996}
Dembo M, Oliver T, Ishihara A, Jacobson K (1996) Imaging the traction stresses
  exerted by locomoting cells with the elastic substratum method.
\newblock Biophys J 70: 2008-22.
\input{dembo1996}

\bibitem{Munevar2001}
Munevar S, li~Wang Y, Dembo M (2001) Traction force microscopy of migrating
  normal and h-ras transformed 3t3 fibroblasts.
\newblock Biophysical Journal 80: 1744 - 1757.
\input{Munevar2001}

\bibitem{butler2002}
Butler JP, Tolic-Norrelykke IM, Fabry B, Fredberg JJ (2002) Traction fields,
  moments, and strain energy that cells exert on their surroundings.
\newblock Am J Physiol Cell Physiol 282: C595-605.
\input{butler2002}

\bibitem{delalamo2007}
del Alamo JC, Meili R, Alonso-Latorre B, Rodriguez-Rodriguez J, Aliseda A,
  et~al. (2007) Spatio-temporal analysis of eukaryotic cell motility by
  improved force cytometry.
\newblock Proc Natl Acad Sci USA 104: 13343-13348.
\input{delalamo2007}

\bibitem{Sabass:2008}
Sabass B, Gardel ML, Waterman CM, Schwarz US (2008) High resolution traction
  force microscopy based on experimental and computational advances.
\newblock Biophysical Journal 94: 207 - 220.
\input{Sabass:2008}

\bibitem{trepat2009physical}
Trepat X, Wasserman M, Angelini T, Millet E, Weitz D, et~al. (2009) {Physical
  forces during collective cell migration}.
\newblock Nature Physics .
\input{trepat2009physical}

\bibitem{Maskarinecetal:09}
Maskarinec SA, Franck C, Tirrell DA, Ravichandran G (2009) Quantifying cellular
  traction forces in three dimensions.
\newblock Proceedings of the National Academy of Sciences .
\input{Maskarinecetal:09}

\bibitem{landau}
Landau LD, Lifshitz EM (1999) Theory of Elasticity.
\newblock Oxford, UK: Butterworth Heinemann, third edition.
\input{landau}

\bibitem{maloney:08}
Maloney JM, Walton EB, Bruce CM, Van~Vliet KJ (2008) Influence of finite
  thickness and stiffness on cellular adhesion-induced deformation of compliant
  substrata.
\newblock Phys Rev E 78: 041923.
\input{maloney:08}

\bibitem{linetal:10}
Lin YC, Tambe DT, Park CY, Wasserman MR, Trepat X, et~al. (2010) Mechanosensing
  of substrate thickness.
\newblock Phys Rev E 82: 041918.
\input{linetal:10}

\bibitem{meilietal:99}
Meili R, Ellsworth C, Lee S, Reddy T, Ma H, et~al. (1999)
  Chemoattractant-mediated transient activation and membrane localization of
  akt/pkb is required for efficient chemotaxis to camp in dictyostelium.
\newblock EMBO J 18: 2092 - 2105.
\input{meilietal:99}

\bibitem{Beningo2002}
Beningo KA, Lo CM, Wang YL (2002) Flexible polyacrylamide substrata for the
  analysis of mechanical interactions at cell-substratum adhesions.
\newblock Academic Press, volume~69 of \emph{Methods in Cell Biology}. pp. 325
  - 339.
\newblock \doi{DOI: 10.1016/S0091-679X(02)69021-1}.
\newblock
  \urlprefix\url{http://www.sciencedirect.com/science/article/pii/S0091679X02690211}.
\input{Beningo2002}

\bibitem{AlonsoLatorre2010}
Alonso-Latorre B (2010) Force and shape coordination in amoeboid cell motility.
\newblock Ph.D. thesis.
\input{AlonsoLatorre2010}

\bibitem{Bastounis11}
Bastounis E, Meili R, Alonso-Latorre B, del \'Alamo JC, Lasheras JC, et~al.
  (2011) Role of the scar/wave complex in regulating traction forces during
  amoeboid motility.
\newblock Molecular Biology of the Cell TBD: TBD.
\input{Bastounis11}

\bibitem{meili2010}
Meili R, Alonso-Latorre B, del Alamo JC, Firtel RA, Lasheras JC (2010) Myosin
  ii is essential for the spatiotemporal organization of traction forces during
  cell motility.
\newblock Molecular Biology of the Cell 21: 405-417.
\input{meili2010}

\bibitem{sung2009}
Hur S, Zhao Y, Li YS, Botvinick E, Chien S (2009) Live cells exert
  3-dimensional traction forces on their substrata.
\newblock Cellular and Molecular Bioengineering 2: 425-436.
\input{sung2009}

\bibitem{chippada2010simultaneous}
Chippada U, Yurke B, Langranaa N (2010) {Simultaneous determination of Young
  modulus, shear modulus, and Poisson ratio of soft hydrogels}.
\newblock J Mater Res 25: 546.
\input{chippada2010simultaneous}

\bibitem{takigawa1996poisson}
Takigawa T, Morino Y, Urayama K, Masuda T (1996) {Poisson ratio of
  polyacrylamide (PAAm) gels}.
\newblock Polymer Gels and Networks 4: 1--5.
\input{takigawa1996poisson}

\bibitem{li1993new}
Li Y, Hu Z, Li C (1993) {New method for measuring Poisson's ratio in polymer
  gels}.
\newblock Journal of Applied Polymer Science 50: 1107--1111.
\input{li1993new}

\bibitem{Legantetal:2010}
Legant WR, Miller JS, Blakely BL, Cohen DM, Genin GM, et~al. (2010) Measurement
  of mechanical tractions exerted by cells in three-dimensional matrices.
\newblock Nat Meth 7: 969--971.
\input{Legantetal:2010}

\bibitem{senetal:09}
Sen S, Engler A, Discher D (2009) Matrix strains induced by cells: Computing
  how far cells can feel.
\newblock Cellular and Molecular Bioengineering 2: 39-48.
\input{senetal:09}

\bibitem{vincentetal:12}
Vincent LG, Yong T, del \'Alamo JC, Tan LP, Engler AJ (2012) Cell aspect ratio
  alters stem cell traction stresses and lineage.
\newblock Biophysical Journal 102: 716a.
\input{vincentetal:12}

\bibitem{Yip201319}
Yip A, Iwasaki K, Ursekar C, Machiyama H, Saxena M, et~al. (2013) Cellular
  response to substrate rigidity is governed by either stress or strain.
\newblock Biophysical Journal 104: 19 - 29.
\input{Yip201319}

\bibitem{Loetal2000}
Lo CM, Wang HB, Dembo M, li~Wang Y (2000) Cell movement is guided by the
  rigidity of the substrate.
\newblock Biophysical Journal 79: 144 - 152.
\input{Loetal2000}

\bibitem{kevcole}
Kevorkian J, Cole JD (2010) Perturbation Methods in Applied Mathematics.
\newblock Applied Mathematical Sciences. Springer Verlag, second edition, 420
  pp.
\input{kevcole}

\end{thebibliography}

\pagebreak
\section*{Figure Legends}

\begin{figure}[!ht]
\begin{center}
\vspace{2ex}
\includegraphics[width=0.99\textwidth]{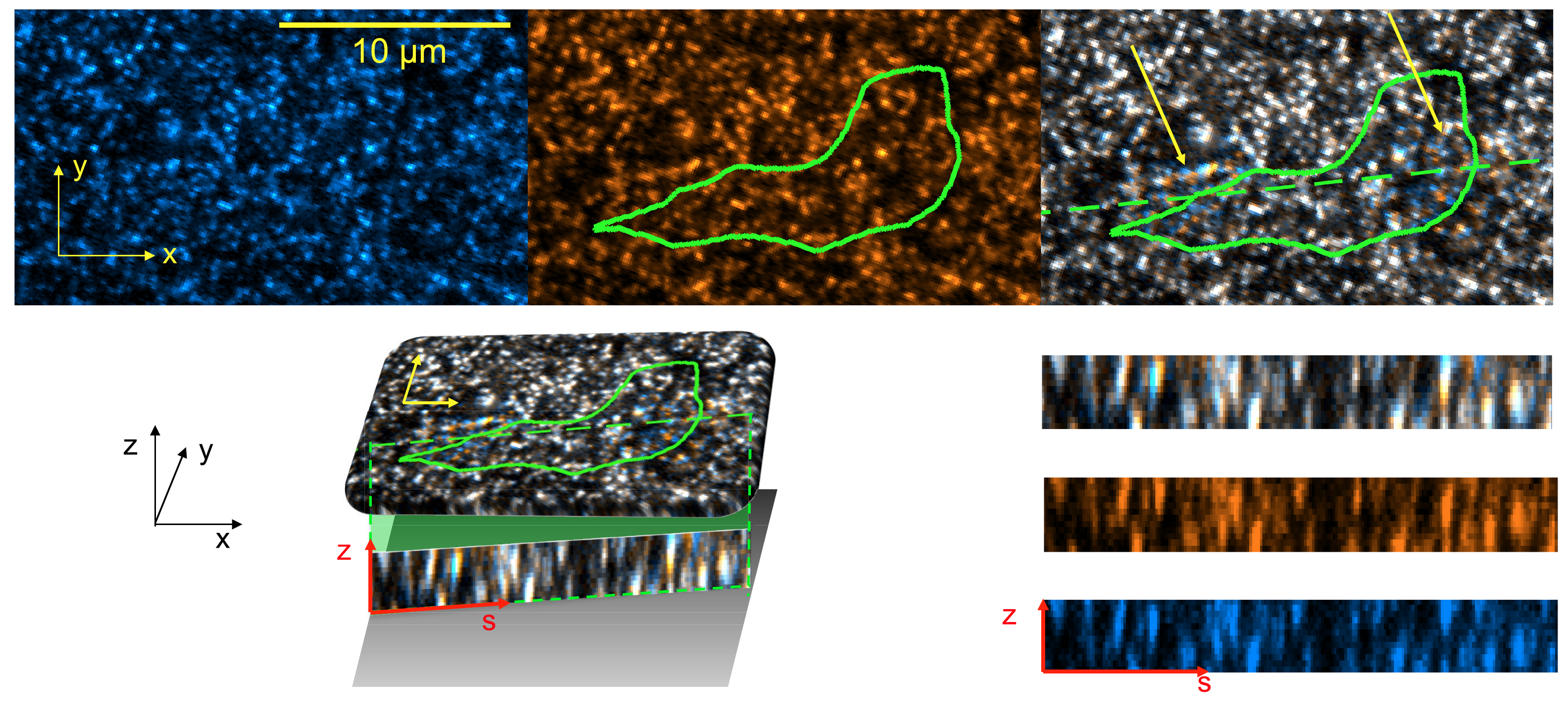}
\mylab{-0.95\textwidth}{0.455\textwidth}{{\footnotesize (\aaa) Undeformed,
horizontal plane}}%
\mylab{-0.64\textwidth}{0.455\textwidth}{{\footnotesize (\bbb) Deformed,
horizontal plane}}%
\mylab{-0.32\textwidth}{0.455\textwidth}{{\footnotesize (\ccc) Merge,
horizontal plane}}%
\mylab{-0.92\textwidth}{0.225\textwidth}{{\footnotesize (\ddd)}}%
\mylab{-0.33\textwidth}{0.234\textwidth}{{\footnotesize (\eee) Merge, vertical
plane}}%
\mylab{-0.33\textwidth}{0.160\textwidth}{{\footnotesize (\fff) Deformed,
vertical plane}}%
\mylab{-0.33\textwidth}{0.085\textwidth}{{\footnotesize (\ggg) Undeformed,
vertical plane}}%
\caption{{\bf Example of fluorescence confocal image used to determine
substratum deformation.}  
The three panels at the top of the figure (\aaa--\ccc) show the first
horizontal slice of a $512\times 512\times 12$--pixel $x-y-z$ stack, which is
focused at the free surface of the substratum ($z=h$).  
\abol, Tracer beads fluorescence in undeformed conditions used as reference for
traction force microscopy.  The scale bar is 10 microns long.  The axes
indicate the reference system for both the substratum deformation and the
traction stresses.
\bbol, Tracer beads fluorescence when the substratum is deformed by a migrating
cell, whose outline is indicated by the green contour.
\cbol, Image obtained by merging the fluorescence from tracer beads in
undeformed (\aaa) and deformed (\bbb) conditions, which reveals the deformation
of the substratum.  White speckles indicate perfect match between undeformed
and deformed conditions and thus zero local deformation.  Blue and orange
speckles indicate mismatch between undeformed and deformed conditions and thus
non-zero local deformation.
Regions of locally large deformation are indicated with yellow arrows.
The dashed green line indicates the location of the vertical section shown in
panels \ddd-\ggg.
\dbol, Three-dimensional illustration of the relative positions of the
horizontal plane in panel \ccc\, ($x-y$, yellow axes) and the vertical plane in
panel \eee\, ($s-z$, red axes).  The black axes indicate the three-dimensional
reference system used to express both the substratum deformation and the
traction stresses. 
The three panels at the bottom right corner of the figure (\eee-\ggg) show
vertical slices of the same $512 \times 512 \times 12$--stack passing through
the dashed green line in panel \ccc\, at $z=h$.
\ebol, Image obtained by merging the fluorescence from tracer beads in
undeformed (\fff) and deformed (\ggg) conditions.  Speckle patterns with orange
top and blue bottom are found in locations where the cell is pulling up on the
substratum.  Conversely, speckle patterns with blue top and orange bottom are
found the when cell is pushing down on the substratum.
\fbol, Tracer beads fluorescence in deformed condition.
\gbol, Tracer beads fluorescence in underformed condition.
}
\label{fig:speckles}
\end{center}
%
\end{figure}

\pagebreak

\begin{figure}[!ht]
\begin{center}
\includegraphics[width=0.8\textwidth]{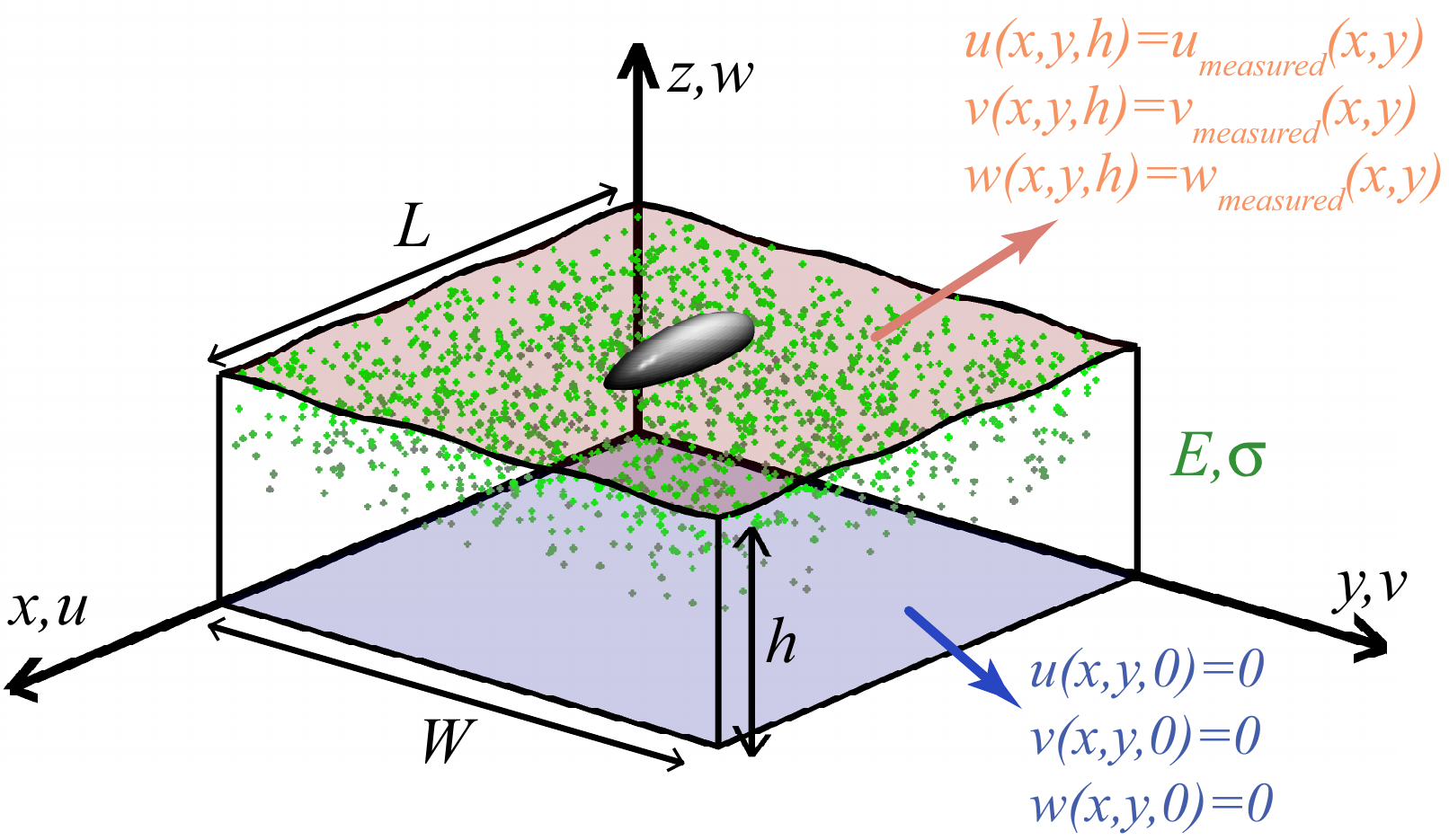}
\caption{
{\bf Configuration of the 3D TFM mathematical problem.}  The input data are the
measured three-dimensional deformation caused by the cell on the free surface
of the substratum ($z=h$, red), and it is assumed that the deformation of the
substratum is zero at the bottom surface in contact with the glass
coverslip ($z=0$, blue).  We assume that the substratum has linear, homogeneous,
isotropic material properties, with Young modulus $E$ and Poisson's ratio
$\sigma$.  Fourier series with spatial periods $L$ and $W$ are used to express
the dependence of the variables in the horizontal directions.
}
\label{fig:sketch}
\end{center}
\end{figure}

\pagebreak

\begin{figure}[!ht]
\begin{center}
\includegraphics[width=0.95\textwidth]{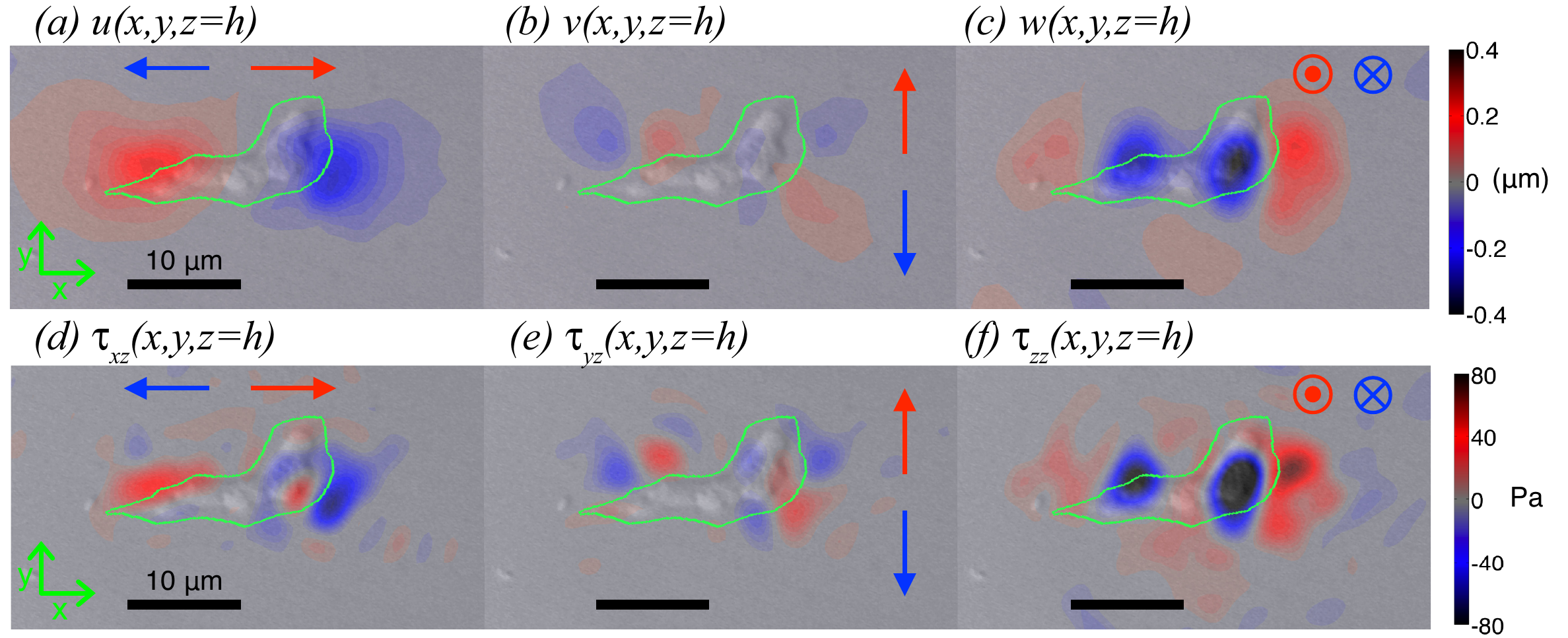}
\caption{
{\bf Three-dimensional deformation field (panels \aaa-\ccc) and stress field
(panels \ddd-\fff) generated by the example cell in figure \ref{fig:speckles}
on the surface of the substratum}.  The cell is moving from left to right. 
The level of deformation / stress is represented by a pseudo-color map
according to the color bars at the right hand side of the figure.  The green
contour indicates the cell outline.  The data are overlaid on the DIC image
used to identify the cell body (see \S \ref{sec:cellcontour}).  The black scale
bars are $10\, \mu m$ long.
\abol, Tangential (horizontal) deformation in the direction parallel to cell
speed, $u(x,y,h)$.
\bbol, Tangential (horizontal) deformation in the direction perpendicular to
cell speed, $v(x,y,h)$.
\cbol, Normal (vertical) deformation, $w(x,y,h)$.
\dbol, Tangential (horizontal) stress in the direction parallel to cell speed,
$\tau_{xz}(x,y,h)$.
\ebol, Tangential (horizontal) stress in the direction perpendicular to cell
speed, $\tau_{yz}(x,y,h)$.
\fbol, Normal (vertical) stress, $\tau_{zz}(x,y,h)$.
The arrows in panels \aaa, \bbb, \ddd\, and \eee\, indicate the directions of
positive (red) and negative (blue) deformation / stress.  The $\otimes$ and
$\odot$ symbols in panels \ccc\, and \fff\, indicate deformation / stress
pointing respectively into the plane (blue, negative) and out of the plane
(red, positive).
}
\label{fig:examplecell}
\end{center}
\end{figure}

\pagebreak
%
\begin{figure}[!ht]
\begin{center}
\vspace{1cm}
\includegraphics[width=0.7\textheight]{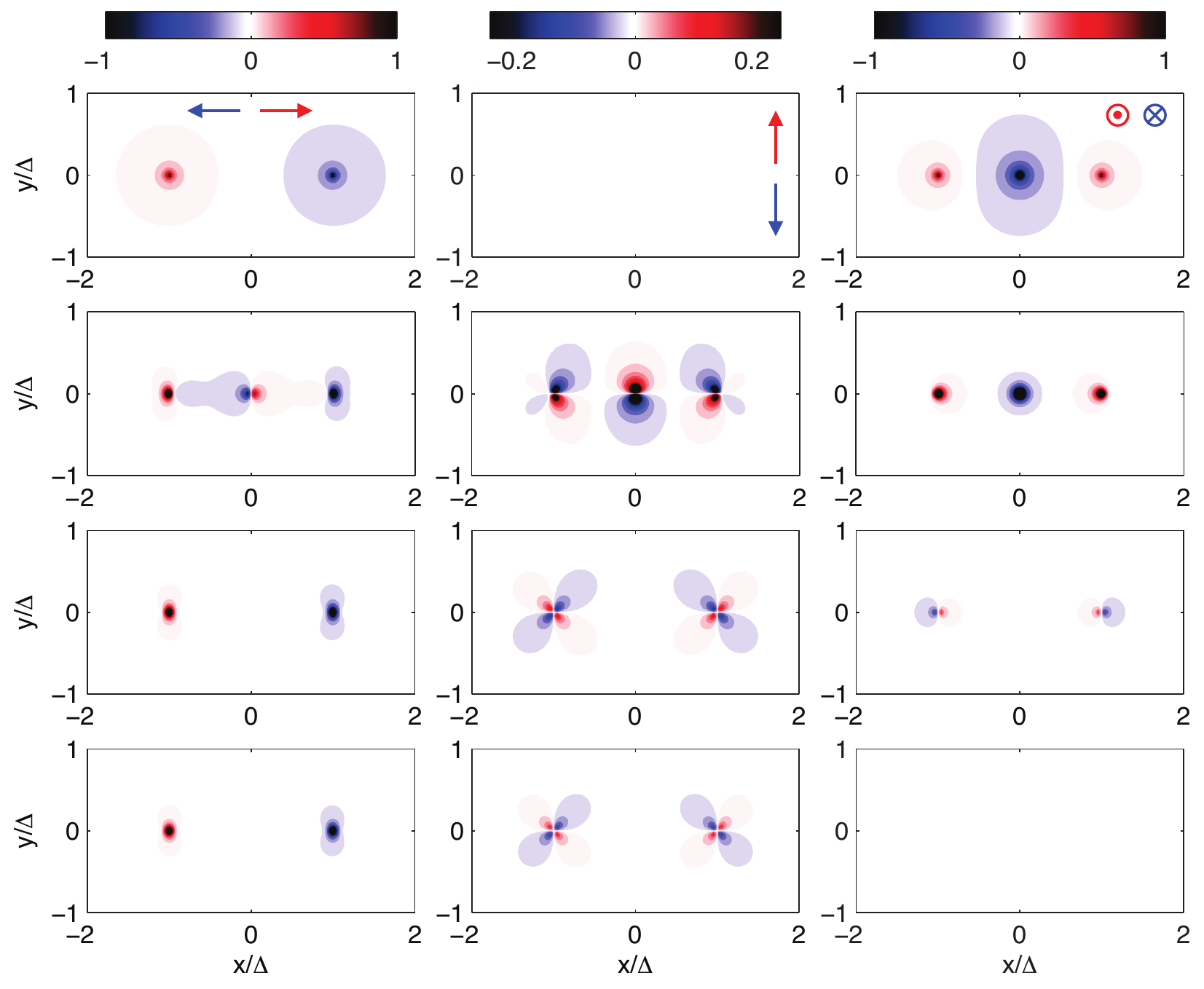}
\mylab{-0.860\textwidth}{0.565\textwidth}{\small (\aaa) $u$}%
\mylab{-0.565\textwidth}{0.565\textwidth}{\small (\bbb) $v$}%
\mylab{-0.273\textwidth}{0.565\textwidth}{\small (\ccc) $w$}%
\mylab{-0.860\textwidth}{0.398\textwidth}{\small (\ddd) $\tau_{xz}$, 3D-TFM}%
\mylab{-0.565\textwidth}{0.398\textwidth}{\small (\eee) $\tau_{yz}$, 3D-TFM}%
\mylab{-0.273\textwidth}{0.398\textwidth}{\small (\fff) $\tau_{zz}$, 3D-TFM}%
\mylab{-0.860\textwidth}{0.233\textwidth}{\small (\ggg) $\tau_{xz}$, 2D-TFM \cite{delalamo2007}}%
\mylab{-0.565\textwidth}{0.233\textwidth}{\small (\hhh) $\tau_{yz}$, 2D-TFM \cite{delalamo2007}}%
\mylab{-0.273\textwidth}{0.233\textwidth}{\small (\iii) $\tau_{zz}$, 2D-TFM \cite{delalamo2007}}%
\mylab{-0.860\textwidth}{0.068\textwidth}{\small (\jjj) $\tau_{xz}$, 2D-TFM \cite{trepat2009physical}}%
\mylab{-0.565\textwidth}{0.068\textwidth}{\small (\kkk) $\tau_{yz}$, 2D-TFM \cite{trepat2009physical}}%
\mylab{-0.273\textwidth}{0.068\textwidth}{\small (\lll) $\tau_{zz}$, 2D-TFM \cite{trepat2009physical}}%
\caption{{\bf Side-by-side comparison of 3D Fourier TFM versus previous 2D
methods \cite{delalamo2007,trepat2009physical} for a synthetic deformation
field representative of the deformation patterns exerted by migrating amoeboid
cells (see figure \ref{fig:examplecell})}.  The Poisson's ratio is $\sigma = 0.3$
and the substratum thickness, $h = 2 \Delta$, is equal to the length of the
``synthetic cell''.
The plots in the top row show the synthetic deformation field in the $x$
direction (eq. \ref{eq:usyn}, panel \aaa),  $y$ direction (zero, panel \bbb)
and $z$ direction (eq. \ref{eq:wsyn}, panel \ccc). 
The second row shows the traction stresses calculated from the displacements in
panels (\aaa)-(\ccc) by 3D Fourier TFM.  (\ddd), $\tau_{xz}$; (\eee),
$\tau_{yz}$; (\fff), $\tau_{zz}$. 
The third row shows the traction stresses calculated from the displacements in
panels (\aaa)-(\ccc) by 2D Fourier TFM under the assumption of zero normal
displacements on the substratum's surface (\ie $w(z=h)=0$ as in ref.
\cite{trepat2009physical}).  (\ggg), $\tau_{xz}$; (\hhh), $\tau_{yz}$; (\iii),
$\tau_{zz}$. 
The last row shows the traction stresses calculated from the displacements in
panels (\aaa)-(\ccc) by 2D Fourier TFM under the assumption of zero normal
stresses on the substratum's surface (\ie $\tau_{zz}(z=h)=0$ as in ref.
\cite{delalamo2007}).  (\jjj), $\tau_{xz}$; (\kkk), $\tau_{yz}$; (\lll),
$\tau_{zz}$. 
}
\label{fig:synthe}
\end{center}
\end{figure}

\pagebreak

\begin{figure}[!ht]

\begin{center}
\vspace{1cm} \includegraphics[width=0.8\textwidth]{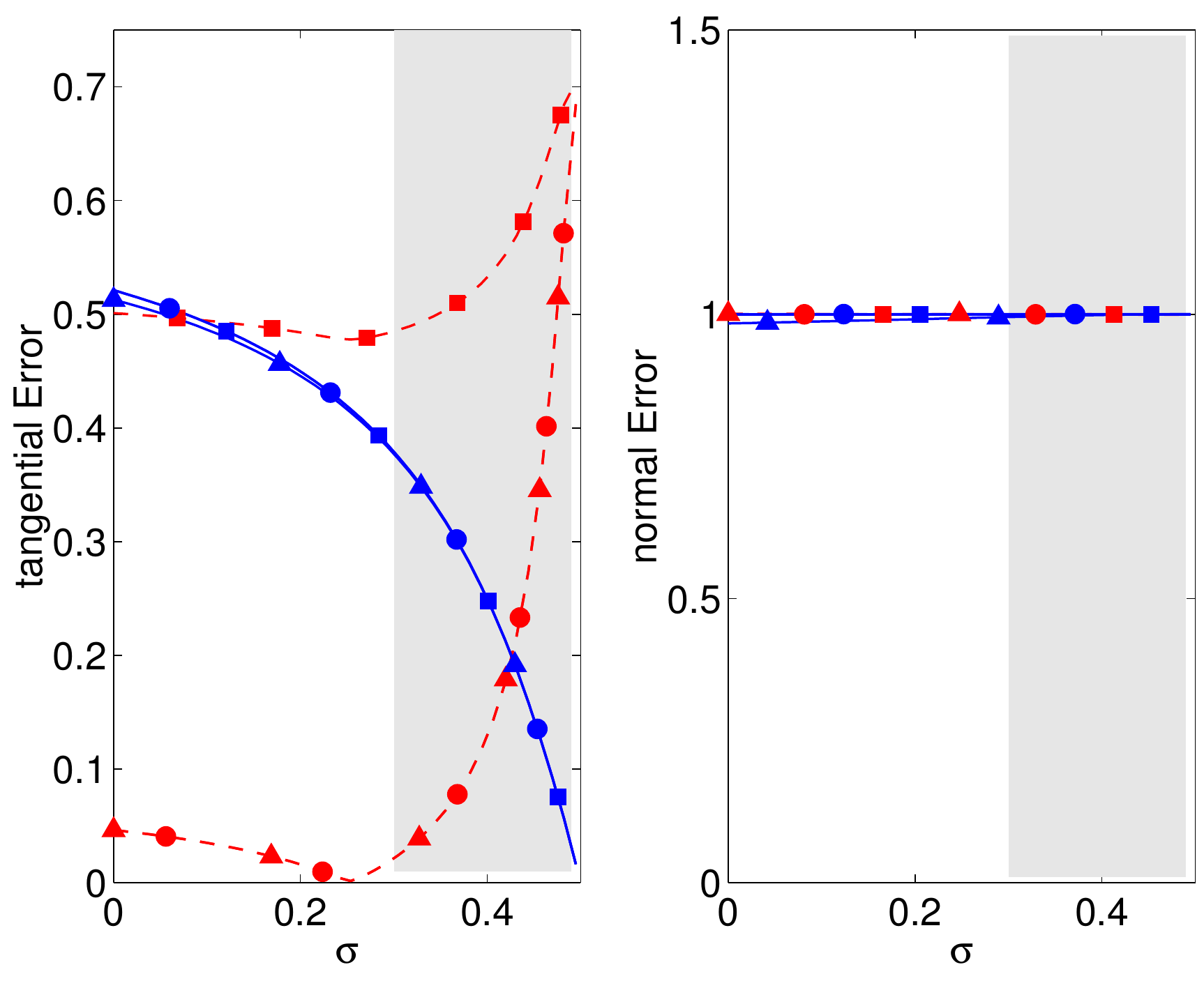}
\mylab{-0.72\textwidth}{0.57\textwidth}{(\aaa)}%
\mylab{-0.31\textwidth}{0.57\textwidth}{(\bbb)}%
\caption{{\bf Relative error of 2D TFM methods represented as a function of the
Poisson's ratio, $\sigma$, for two values of the ratio $\Delta/h$.}  
Red lines and symbols, $\Delta/h=100$; blue lines and symbols, $\Delta/h =
0.01$.
\linesolidcircle, 2D method with finite $h$ and $\tau_{zz}=0$ on the surface
(ref. \cite{delalamo2007}); \linesolidtrian, 2D method with finite $h$ and
$w=0$ on the surface (ref. \cite{trepat2009physical}); \linesolidsquar,
Boussinesq solution with infinite $h$ (refs. \cite{dembo1996,butler2002}).
(\aaa), $E_t$; (\bbb), $E_n$.  The shaded patch represents the range of values
of Poisson's ratio reported for gels customarily employed in TFM
\cite{chippada2010simultaneous,takigawa1996poisson,li1993new}.
}
%
\label{fig:errsig1D}
\end{center}
\end{figure}

\pagebreak

\begin{figure}[!ht]

\begin{center}
\vspace{1cm} \includegraphics[width=0.8\textwidth]{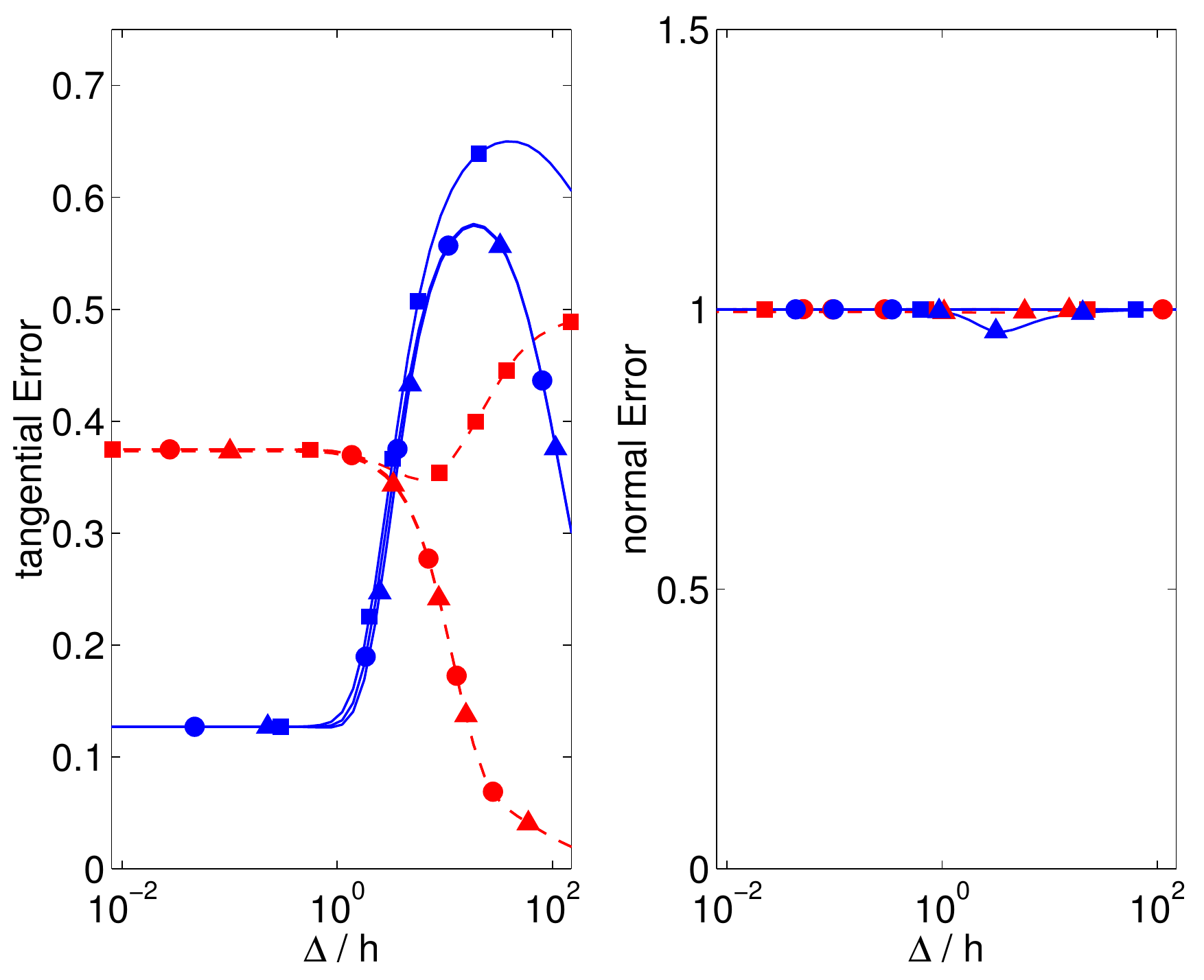}
\mylab{-0.72\textwidth}{0.57\textwidth}{(\aaa)}%
\mylab{-0.31\textwidth}{0.57\textwidth}{(\bbb)}%
\caption{{\bf Relative error of 2D TFM represented as a function of $\Delta/ h$
for two values of the Poisson's ratio representative of polyacrylamide gels.}
Red lines and symbols, $\sigma = 0.3$; blue lines and symbols, $\sigma= 0.45$.
\linesolidcircle, 2D method with finite $h$ and $\tau_{zz}=0$ on the surface
(ref. \cite{delalamo2007}); \linesolidtrian, 2D method with finite $h$ and
$w=0$ on the surface (ref. \cite{trepat2009physical}); \linesolidsquar,
Boussinesq solution with infinite $h$ (refs.
\cite{dembo1996,butler2002}).
}
\label{fig:errhh1D}
\end{center}
\end{figure}

\pagebreak

\begin{figure}[!ht]
\begin{center}
\vspace{1cm} \includegraphics[width=0.95\textwidth]{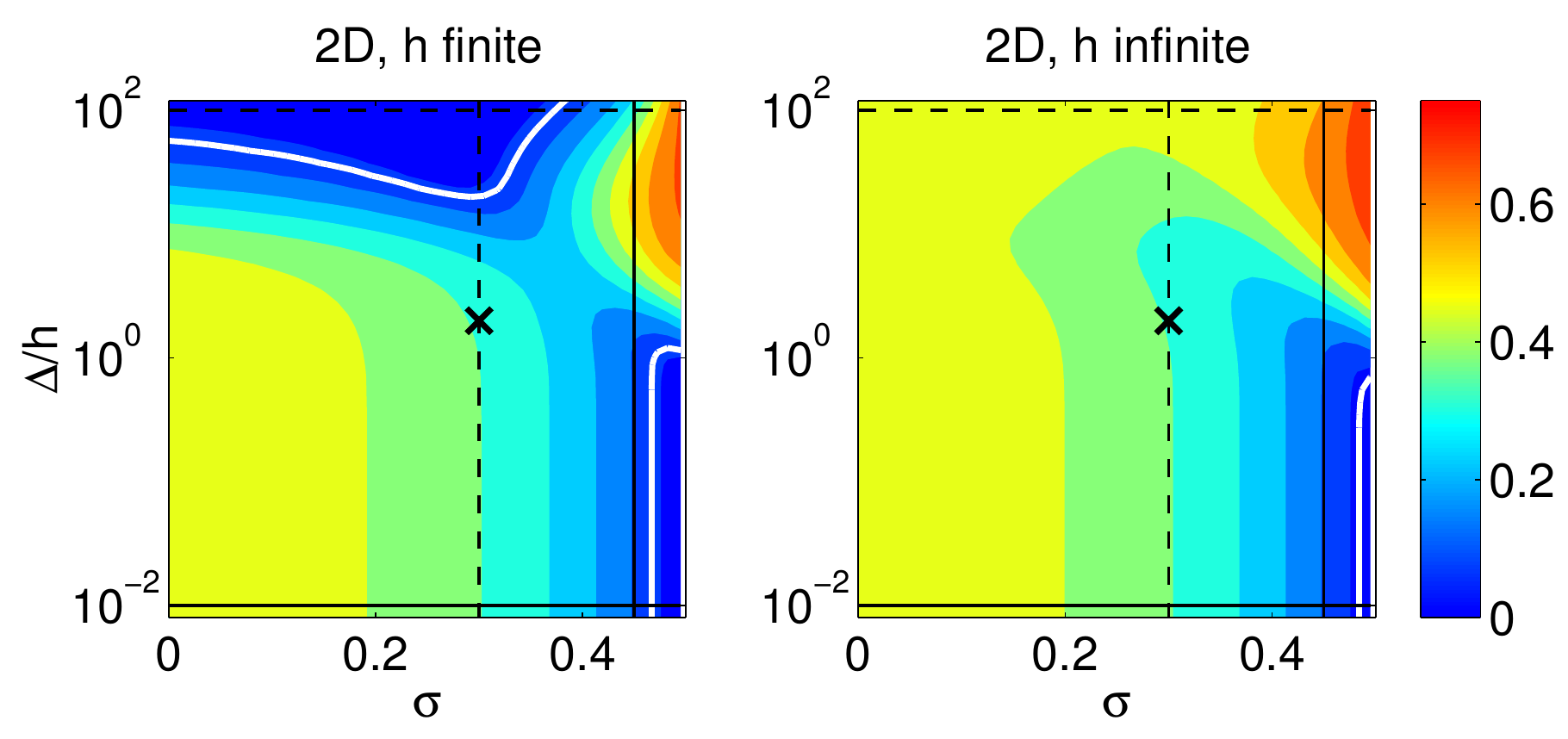}
\mylab{-0.85\textwidth}{0.42\textwidth}{(\aaa)}%
\mylab{-0.42\textwidth}{0.42\textwidth}{(\bbb)}%
\caption{
{\bf Relative error of 2D TFM represented as a function of the Poisson's ratio,
$\sigma$, and the ratio $\Delta/h$}. (\aaa), $\mathcal{E}_t$ assuming zero
normal deformation on the surface of the substratum; (\bbb), $\mathcal{E}_t$
from Boussinesq's solution.  The black $\times$ marks the combination of
$\sigma$ and $\Delta/h$ used to calculate the traction stress maps in figure
\ref{fig:synthe}.  The horizontal lines mark the values of $\Delta/h$ used to
plot figure \ref{fig:errsig1D}.  \solid, $\Delta/h=0.01$; \dashed,
$\Delta/h=10$.
The thick white contours correspond to $\mathcal{E}_t = 0.1$. 
The vertical lines mark the values of $\sigma$ used to plot figure
\ref{fig:errhh1D}. \dashed, $\sigma=0.3$; \solid, $\sigma=0.45$.
}
%
\label{fig:err2Dsig2D}
\end{center}
\end{figure}

\pagebreak
%
\begin{figure}[!ht]
\begin{center}
\vspace{1cm} \includegraphics[width=0.95\textwidth]{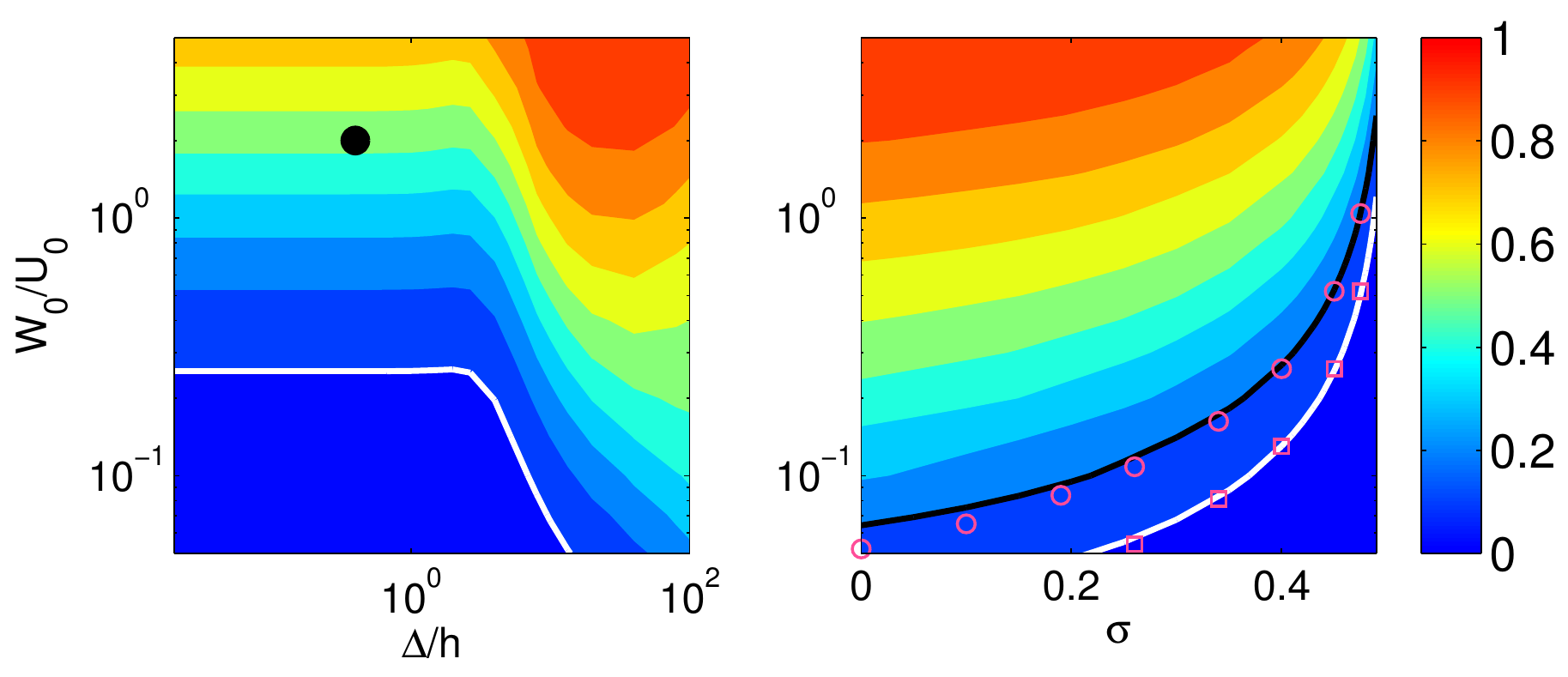}
\mylab{-0.85\textwidth}{0.42\textwidth}{(\aaa)}%
\mylab{-0.42\textwidth}{0.42\textwidth}{(\bbb)}%
\caption{{\bf Effect of the ratio of normal to horizontal deformation,
$W_0/U_0$ on the error of 2D TFM.}
(\aaa), contour map of $\mathcal{E}_t$ for $\sigma = 0.45$, represented as a
function of $\Delta / h$ and $W_0/U_0$.  The black circle corresponds to the
example cell in figure \ref{fig:examplecell}.  
The thick white contour corresponds to $\mathcal{E}_t = 0.1$. 
(\bbb), contour map of $\mathcal{E}_t$ for $\Delta = h/2$, represented as a
function of $\sigma$ and $W_0/U_0$.  
The thick white (black) contours correspond to $\mathcal{E}_t = 0.1$ ($0.2$),
and are well approximated by the magenta squares (circles) coming the eq.
\ref{eq:errform}}.
%
\label{fig:errW}
\end{center}
\end{figure}

\pagebreak

\begin{figure}[!ht]
\begin{center}
\includegraphics[width=0.6\textwidth]{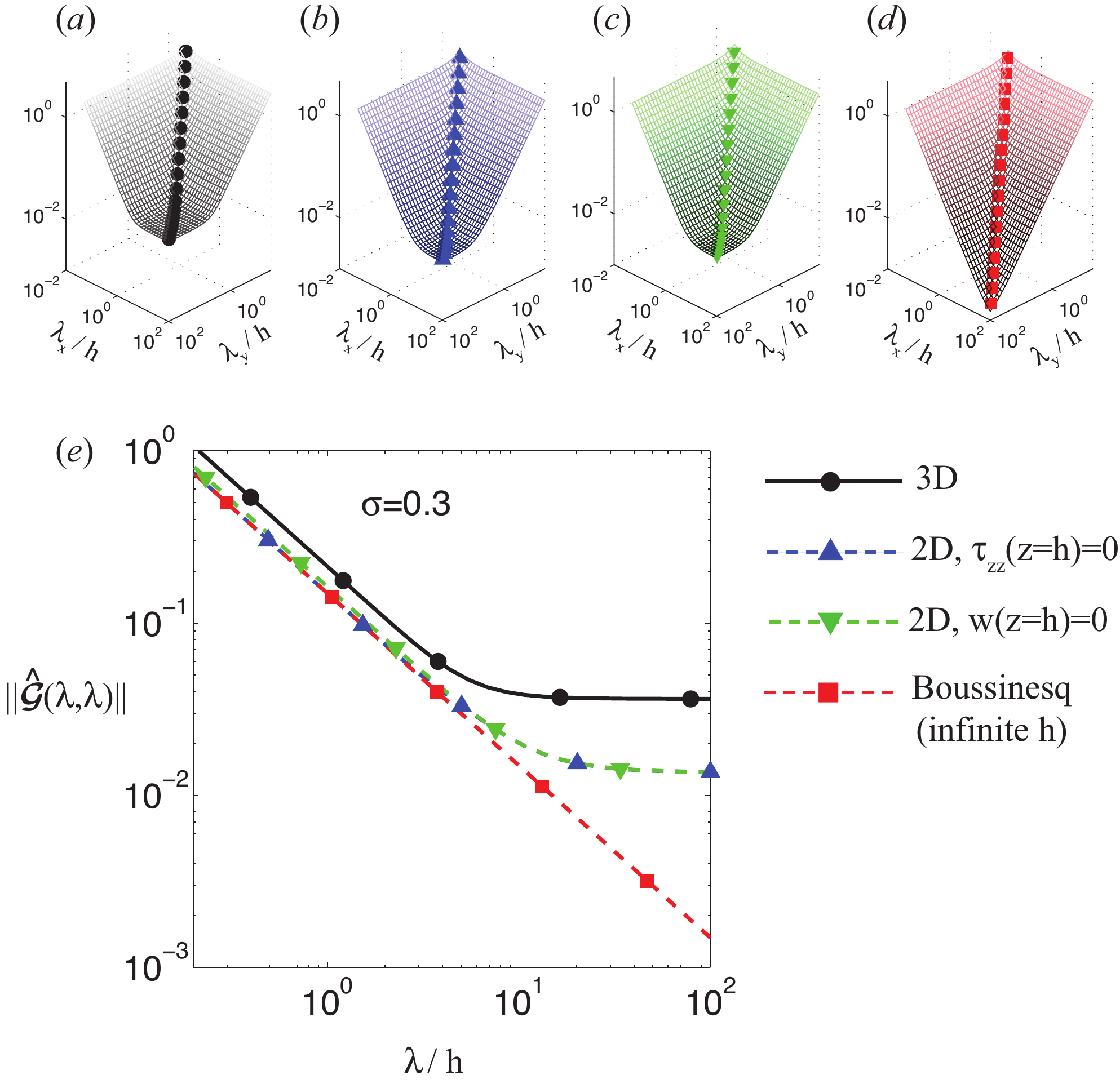}
\caption{{\bf Fr\"obenius norm of the Green's function used by different TFM
methods, $||{\bf \mathcal{\widehat G}}||$ (eq. \ref{eq:norma}), for $\sigma =
0.3$}.  The four panels in the top row (\aaa--\ddd) show surface plots of
$||{\bf \mathcal{\widehat G}}||$ as a function of the horizontal wavelengths of
the strain / stress fields $(\lambda_x,\lambda_y)$.  \abol, present 3D TFM
method; 
\bbol, 2D TFM under the assumption of zero normal stresses on the substratum's
surface (\ie $\tau_{zz}(z=h)=0$ as in ref.  \cite{delalamo2007}); 
\cbol, 2D TFM under the assumption of zero normal displacements on the
substratum's surface (\ie $w(z=h)=0$ as in ref.  \cite{trepat2009physical});
\dbol, Boussinesq's traction cytometry assuming an infinitely-thick substratum
(as in refs.  \cite{dembo1996,butler2002}).  The symbol curves in these plots
indicate the sections of $||{\bf \mathcal{\widehat G}}||$ represented in panel
(\eee).  \ebol,  $||{\bf \mathcal{\widehat G}}||$ along the line $\lambda =
\lambda_x = \lambda_y$ from different traction TFM methods, represented as a
function of $\lambda / h$.
}
\label{fig:kerncomp}
\end{center}
\end{figure}

\pagebreak

\begin{figure}[h!]
\includegraphics[width=0.95\textwidth]{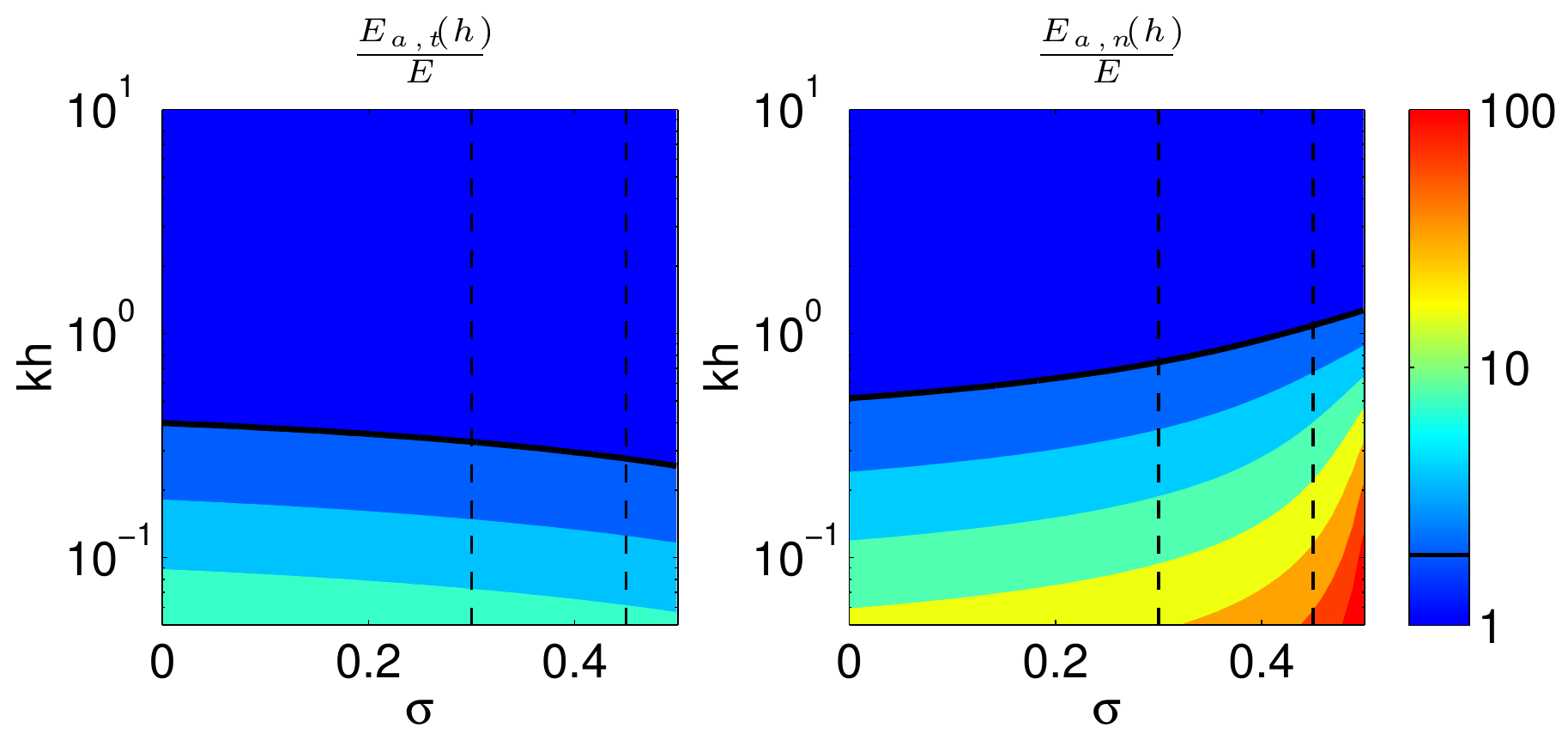}
\mylab{-0.87\textwidth}{0.46\textwidth}{(\aaa)}%
\mylab{-0.44\textwidth}{0.46\textwidth}{(\bbb)}%
\caption{
{\bf
Effect of finite substratum thickness on the substratum stiffness mechanosensed
by tangential or normal traction stresses.
}
Panels (\aaa) and (\bbb) display the apparent elastic modulus of the substratum
(eq. \ref{eq:apparentE}) in the directions tangential and normal to the surface
respectively. 
The data are plotted as a function of the Poisson's ratio, $\sigma$, and the
substratum thickness $h$ normalized with the wavenumber $k = \sqrt{\alpha^2 +
\beta^2}$ of the strain/stress fields.  For simplicity, the $\alpha = \beta$
case is represented but similar results are obtained for other combinations of
the wavenumbers.
The isolines plotted in panels (\aaa) and (\bbb) are respectively $1 \dots
(\times 2) \dots 8$, and $1 \dots (\times 2) \dots 128$.
Particularly, the thick black contour in each panel represents the isoline $E_{a} = 2E$
corresponding to a two-fold increase in apparent stiffness.
The vertical dashed lines indicate the values of the Poisson's ratio
represented in figure \ref{fig:deepH}, $\sigma =0.3, \, 0.45$.
}
\label{fig:deep2D}
\end{figure}

\pagebreak

\begin{figure}[h!]
\includegraphics[height=0.5\textwidth]{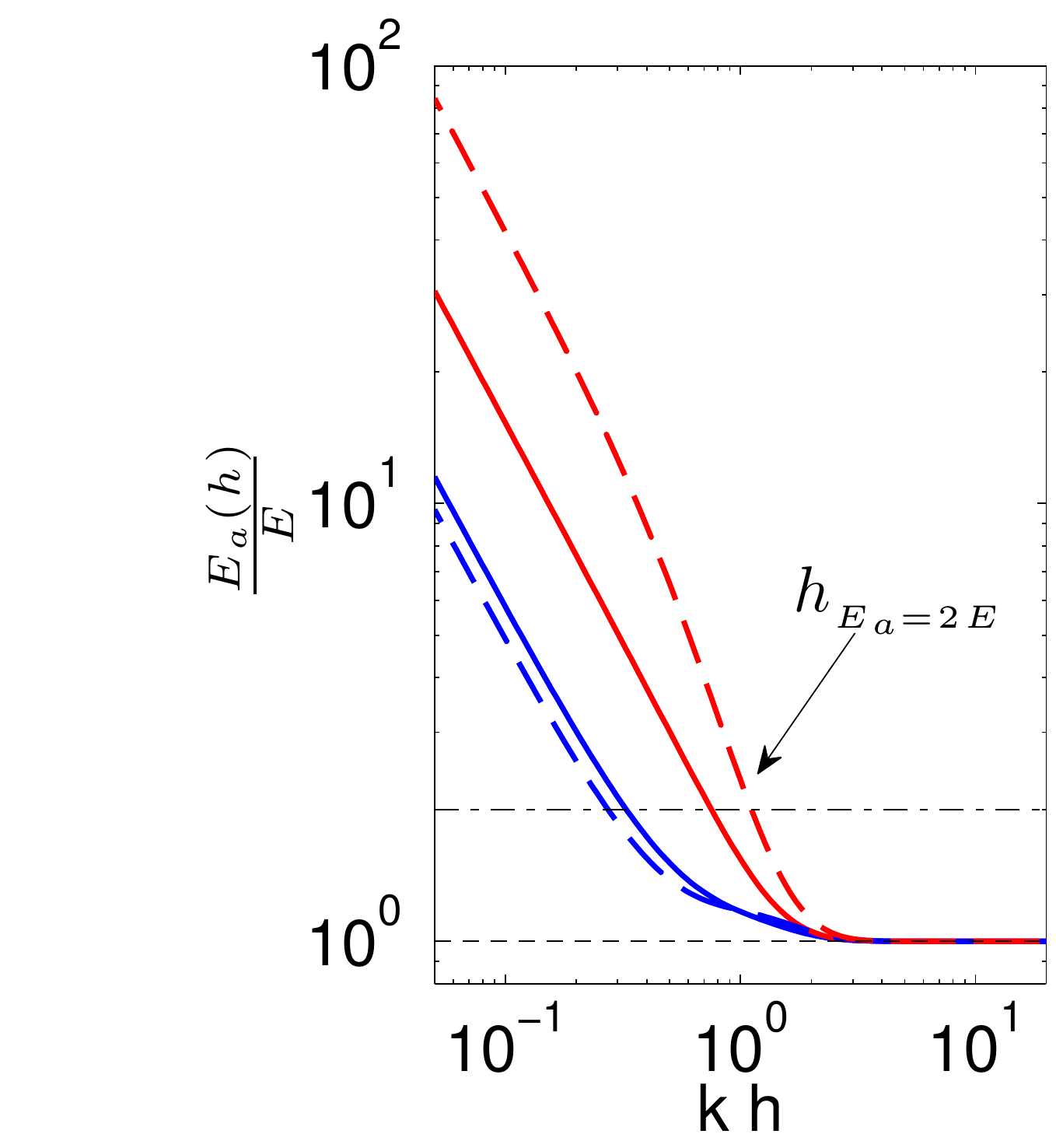}
\includegraphics[height=0.5\textwidth]{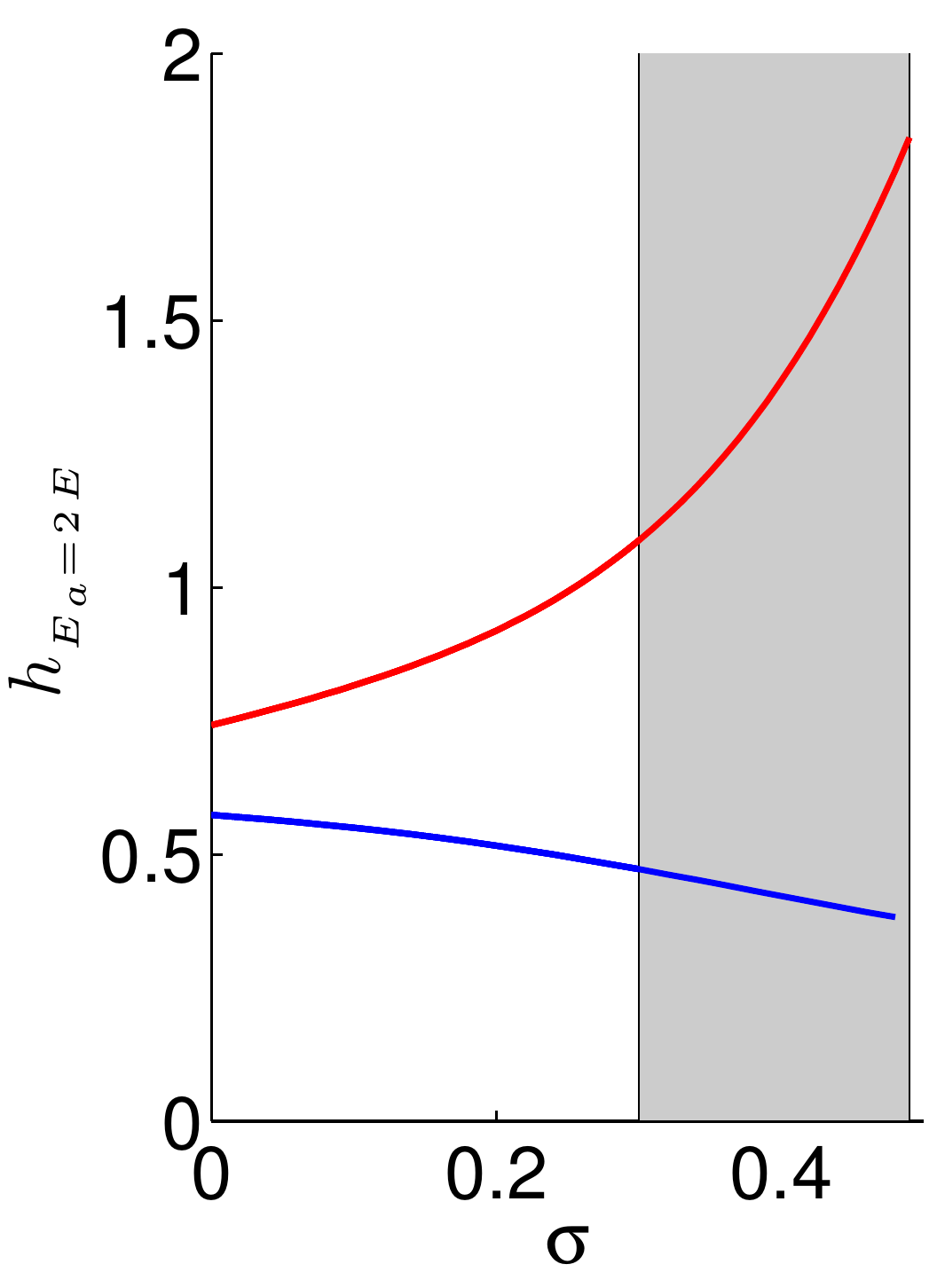}
\mylab{-0.67\textwidth}{0.5\textwidth}{(\aaa)}%
\mylab{-0.34\textwidth}{0.5\textwidth}{(\bbb)}%
\caption{
{\bf Sensing depth by tangential and normal traction stresses.}
Panel \abol \, displays the ratio $E_a/E$ as a function of the substratum
thickness $h$ normalized with the inverse wavenumber $k = \sqrt{\alpha^2 +
\beta^2}$ of the strain/stress fields.  For simplicity, the $\alpha = \beta$
case is represented but similar results are obtained for other combinations of
the wavenumbers.
\rojo{\solid}, normal direction and $\sigma = 0.3$; \azul{\solid}, tangential
direction and $\sigma=0.3$; \rojo{\dashed}, normal direction and $\sigma =
0.45$; \azul{\dashed}, tangential direction and $\sigma=0.45$.  
The black horizontal lines indicate the levels $E_a = E$ (no increase in
apparent elastic modulus, \dashed) and $E_a = 2E$ (two-fold increase, \chndot).
Panel \bbol \, displays the sensing depth  defined as the value of $h$ that
yields a two-fold increase in apparent elastic modulus compared to $h =
\infty$.
The sensing depth is represented as a function of the Poisson's ratio for
tangential and normal traction stresses.
\rojo{\solid}, normal direction; \azul{\solid}, tangential direction.
The shaded patch represents the range of values of Poisson's ratio measured for
polymer networks \cite{chippada2010simultaneous,takigawa1996poisson,li1993new}
}
\label{fig:deepH}
\end{figure}
\pagebreak
%
\begin{figure}[!ht]
\begin{center}
\vspace{1cm}
\includegraphics[width=0.9\textwidth]{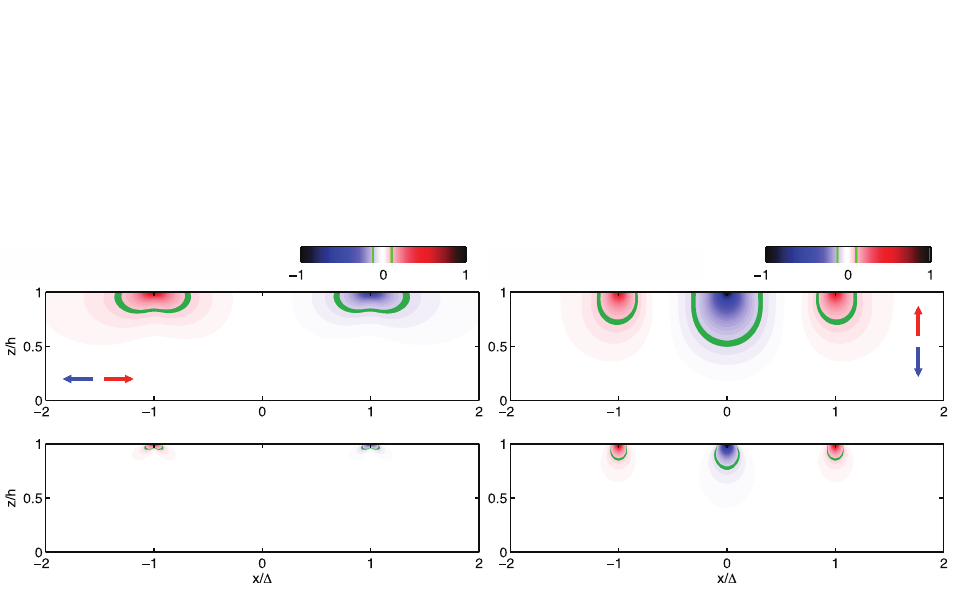}
\mylab{-0.860\textwidth}{0.31\textwidth}{\small (\aaa) $u,\, v(h)=w(h)=0$}%
\mylab{-0.41\textwidth}{0.31\textwidth}{\small (\bbb) $w,\, u(h)=v(h)=0$}%
\mylab{-0.860\textwidth}{0.05\textwidth}{\small (\ddd) $\tau_{xz}$,
$v(h)=w(h)=0$}%
\mylab{-0.41\textwidth}{0.05\textwidth}{\small (\eee) $\tau_{zz}$,
$u(h)=v(h)=0$}%
\caption{{\bf Penetration of tangential and normal deformations and
stresses into the substratum.}
(\aaa) Vertical contour map of $u$ obtained by applying a unit tangential
synthetic deformation field at the free surface of the gel (eqs.
\ref{eq:usyn}-\ref{eq:wsyn} with $U_0=1$ and $W_0=0$), and solving the
elastostatic equation for different values of $z$.
The deformation is plotted in the normal plane $y=0$ as a function of
$x/\Delta$ and $z/h$.  Thus, $z=0$ represents the bottom of the gel in contact
with the coverslip and $z=h$ represents the free surface of the gel.
(\bbb) Same as (\aaa) for the normal deformation $w$ obtained by applying a
unit normal deformation at the gel surface (eqs.  \ref{eq:usyn}-\ref{eq:wsyn}
with $U_0=0$ and $W_0=1$).
(\ccc) Same as (\aaa) for $\tau_{xz}$.
(\ddd) Same as (\bbb) for $\tau_{zz}$.
In all panels, the data are normalized between $-1$ and $1$, and the green
contour represents the $10\%$ iso-level. 
}
\label{fig:penetra}
\end{center}
\end{figure}


\pagebreak

\section*{Tables}

\begin{table*}[!ht]
\begin{center}
\begin{tabular}{ccccc}
\hline \\[-2ex]
{\small Interrogation Box Size }  & {\small Nyquist resolution } & {\small
Signal-to-noise Ratio}  &
\multicolumn{2}{c}{
 \small Displacement floor}
\\
{\small ($\Delta x \times \Delta y$, pixels)}  & {\small ($\mu$m)}  & & {\small $u-v$ ($\mu$m)} & {\small $w$ ($\mu$m)}
\\ \hline
{\small $8 \times 8$}  &  $0.7$  &{\small $1.8$} & {\small $0.014$} & {\small $0.015$} 
\\
{\small $16 \times 16 $} & $1.4$  & {\small $2.3$} & {\small $0.009$} & {\small $0.008$} 
\\
{\small $24 \times 24 $} & $2.1$  & {\small $2.6$} & {\small $0.007$} & {\small $0.006$} 
\\
{\small $32 \times 32 $} & $2.7$  & {\small $2.9$} & {\small $0.007$} & {\small $0.006$} 
\\
{\small $64 \times 64 $} & $5.6$ & {\small $3.5$} & {\small $0.007$} & {\small $0.005$} 
\\ \hline
\vspace{-1cm}
\end{tabular}
\end{center}
\caption{Average signal-to-noise ratio of the three-dimensional displacement
measurements obtained by image cross-correlation as a function of the size of
the interrogation box (in pixels).  Signal-to-noise ratio is defined as the
ratio between the maximum value of the cross-correlation and the second local
maximum.  All data in this table have been obtained for interrogation boxes
with vertical size of $\Delta z=8$ pixels.}
\label{tab:s2n}
\end{table*}

\pagebreak


\section*{Supporting Information}

This appendix provides the explicit mathematical expressions to be plugged in
the equations presented in \S \ref{sec:calcu} to determine the traction
stresses.  These formulae should be preferred to those in the Supplementary
Information of del \'Alamo \etal \cite{delalamo2007}.  The latter were
particularized to two-dimensional boundary conditions and contained a number of
typographical errors that have been corrected here.  The resolvant matrix of
the Fourier transform of the elastostatic equation is

\begin{equation}
{\bf U}(\alpha,\beta,z)=
\left[\begin{array}{c} 
{\frac{{ \alpha}^{2}z\cosh\left(k\, z\right)}{4{k}^{2}\left(1-{
\sigma}\right)}}
+{\frac{ 
[4(1-\sigma)k^2 - \alpha^2]
\sinh\left(k\, z\right)}{4{k}^{3}\left(1-{ \sigma}\right)}}
\\ 
\noalign{\medskip}\,
{\frac{{ \alpha}\,{ \beta}\, z\cosh\left(k\, z\right)}{4{k}^{2}\left(1-
\sigma\right)}}
-\frac{ \alpha \,{ \beta}\,\sinh\left(k\, z\right)}
{4{k}^3(1- \sigma)}
\\ 
\noalign{\medskip}{
\frac{- i \sinh\left(k\, z\right)z{ \alpha}}{4k(1- \sigma)}} 
\end{array}\right]
%
\end{equation}

\begin{equation}
{\bf V}(\alpha,\beta,z)=
\left[\begin{array}{c} 
\noalign{\medskip}\,
{\frac{{ \alpha}\,{ \beta}\, z\cosh\left(k\, z\right)}{4{k}^{2}\left(1-
\sigma\right)}}
-\frac{ \alpha \,{ \beta}\,\sinh\left(k\, z\right)}
{4{k}^3(1- \sigma)}
\\
{\frac{{ \beta}^{2}z\cosh\left(k\, z\right)}{4{k}^{2}\left(1-{
\sigma}\right)}}
+{\frac{ 
[4(1-\sigma)k^2 - \beta^2]
\sinh\left(k\, z\right)}{4{k}^{3}\left(1-{ \sigma}\right)}}
\\ 
\noalign{\medskip}{
\frac{- i \sinh\left(k\, z\right)z{ \beta}}{4k(1- \sigma)}} 
\end{array}\right]
\end{equation}

\begin{equation}
%
{\bf W}(\alpha,\beta,z)=
\left[\begin{array}{c} 
{\frac{ -i \alpha\, z\sinh\left(k\, z\right)}{2k\,\left(1-2\, \sigma\right)}}
\\ 
\noalign{\medskip}{
\frac{-i \beta\, z\sinh\left(k\, z\right)}{2k\,\left(1-2\, \sigma\right)}}
\\
\noalign{\medskip}
\frac{-z\cosh\left(k\, z\right)}{2(1-2\, \sigma)}
+{\frac{\left(3-4\, \sigma\right)\sinh\left(k\, z\right)}{2k\,\left(1-2\,
\sigma\right)}} 
\end{array}\right].
%
\end{equation}

The inverse of the resolvant matrix particularized at the surface of the substratum is given by
\begin{equation}
%
{\left[\begin{array}{c|c|c} 
{\bf U}_{mn}(h)  & {\bf V}_{mn}(h) & {\bf W}_{mn}(h) 
\end{array}\right]}^{-1}=
\left[\begin{array}{ccc} C_{1u} & C_{1v} & C_{1w}\\ C_{2u} & C_{2v} & C_{2w}\\
C_{3u} & C_{3v} & C_{3w}
\end{array}\right],
%
%
\end{equation}
where
\begin{equation*}
%
{\textstyle
C_{1u}=\frac{\left(-4\,{\it \beta}^{2}{h}^{2}+2\,\left(-3+4\,{\it
\sigma}\right)^{2}\left(\cosh\left(2\, k\,
h\right)-1\right)\right){k}^{5}+8\,{\it \alpha}^{2}h\,\left(-1+{\it
\sigma}\right)\sinh\left(2\, k\, h\right){k}^{4}-2\,{\it
\alpha}^{2}\left(\cosh\left(2\, k\, h\right)-1\right)\left(-3+4\,{\it
\sigma}\right){k}^{3}}{{k}^{4}\left(4\,{h}^{2}{k}^{2}+3\,\left(-3+4\,{\it
\sigma}\right)^{2}\right)\sinh\left(k\, h\right)-\left(-3+4\,{\it
\sigma}\right)^{2}{k}^{4}\sinh\left(3\, k\, h\right)}},
%
\end{equation*}
%
\begin{equation*}
%
{\textstyle
C_{1v}=\frac{4\,{\it \alpha}\,{k}^{5}{h}^{2}{\it \beta}+8\,{\it \alpha}\,{\it
\beta}\, h\,\left(-1+{\it \sigma}\right)\sinh\left(2\, k\,
h\right){k}^{4}+\left(-2\,{\it \alpha}\,{\it \beta}\,\left(-3+4\,{\it
\sigma}\right)\cosh\left(2\, k\, h\right)+2\,{\it \alpha}\,{\it
\beta}\,\left(-3+4\,{\it
\sigma}\right)\right){k}^{3}}{{k}^{4}\left(4\,{h}^{2}{k}^{2}+3\,\left(-3+4\,{\it
\sigma}\right)^{2}\right)\sinh\left(k\, h\right)-\left(-3+4\,{\it
\sigma}\right)^{2}{k}^{4}\sinh\left(3\, k\, h\right)} 
},
%
\end{equation*}
%
\begin{equation*}
%
{\textstyle
C_{1w}=\frac{-8\, i{k}^{5}{\it \alpha}\,\left(-1+{\it
\sigma}\right)h\,\left(\cosh\left(2\, k\,
h\right)-1\right)}{{k}^{4}\left(4\,{h}^{2}{k}^{2}+3\,\left(-3+4\,{\it
\sigma}\right)^{2}\right)\sinh\left(k\, h\right)-\left(-3+4\,{\it
\sigma}\right)^{2}{k}^{4}\sinh\left(3\, k\, h\right)}
},
%
\end{equation*}
%
\begin{equation*}
%
C_{2u}(\alpha,\beta)=C_{1v}(\beta,\alpha),
\end{equation*}
%
\begin{equation*}
C_{2v}(\alpha,\beta)=C_{1u}(\beta,\alpha),
\end{equation*}
%
\begin{equation*}
C_{2w}(\alpha,\beta)=C_{1w}(\beta,\alpha),
\end{equation*}
%
\begin{equation*}
%
{\textstyle
C_{3u}(\alpha,\beta)=\frac{-4\, i{\it \alpha}\,\left(-1+2\,{\it
\sigma}\right)h\,\sinh\left(k\, h\right){k}^{3}}{\left(-3+4\,{\it
\sigma}\right)^{2}{k}^{2}\cosh\left(2\, k\,
h\right)-2\,{h}^{2}{k}^{4}-\left(-3+4\,{\it \sigma}\right)^{2}{k}^{2}}
},
%
\end{equation*}
%
\begin{equation*}
C_{3v}(\alpha,\beta)=C_{3u}(\beta,\alpha),
\end{equation*}
%
\begin{equation*}
%
{\textstyle
C_{3w}\frac{-4\,{k}^{4}h\,\left(-1+2\,{\it \sigma}\right)\cosh\left(k\,
h\right)+4\,\sinh\left(k\, h\right)\left(-1+2\,{\it
\sigma}\right)\left(-3+4\,{\it \sigma}\right){k}^{3}}{\left(-3+4\,{\it
\sigma}\right)^{2}{k}^{2}\cosh\left(2\, k\,
h\right)-2\,{h}^{2}{k}^{4}-\left(-3+4\,{\it \sigma}\right)^{2}{k}^{2}}
}.
\end{equation*}

Finally, the linear operator that defines Hooke's law in Fourier space can be written as
\begin{equation}
%
\mathcal{H}=
\frac{E}{2(1+\sigma)}
\left[\begin{array}{cccccc}
0 & 0 & i\alpha_m & 1 & 0 & 0\\
0 & 0 & i\beta_n & 0 & 1 & 0\\
\frac{2i\alpha_m\sigma}{(1-2\sigma)} & \frac{2i\beta_n\sigma}{(1-2\sigma)} & 0 & 0 & 0 & \frac{2(1-\sigma)}{(1-2\sigma)}
\end{array}\right].
\end{equation}


\begin{figure}[!ht]
\begin{center}
\vspace{1cm}
\includegraphics[width=0.7\textheight]{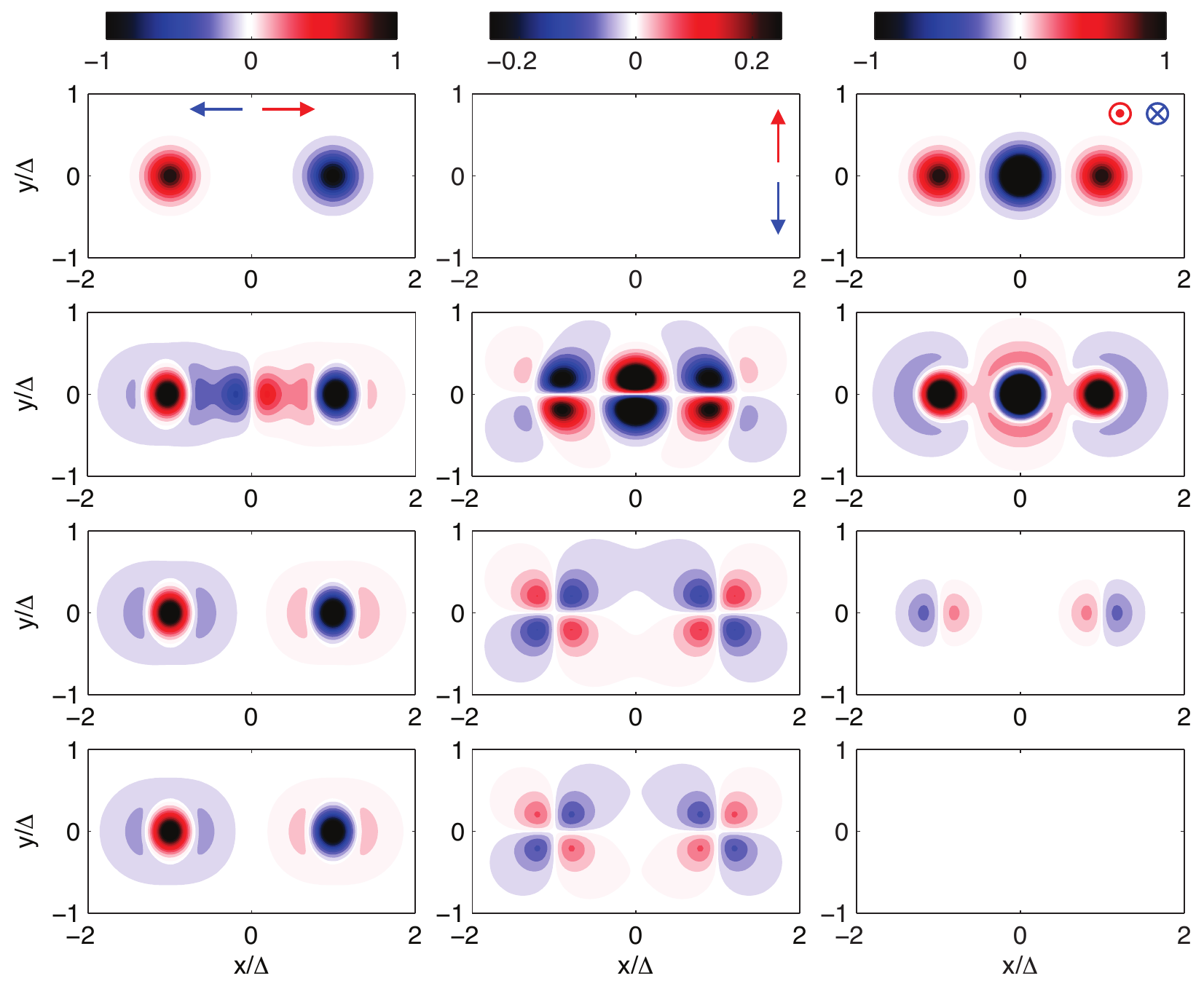}
\mylab{-0.860\textwidth}{0.565\textwidth}{\small (\aaa) $u$}%
\mylab{-0.565\textwidth}{0.565\textwidth}{\small (\bbb) $v$}%
\mylab{-0.273\textwidth}{0.565\textwidth}{\small (\ccc) $w$}%
\mylab{-0.860\textwidth}{0.398\textwidth}{\small (\ddd) $\tau_{xz}$, 3D-TFM}%
\mylab{-0.565\textwidth}{0.398\textwidth}{\small (\eee) $\tau_{yz}$, 3D-TFM}%
\mylab{-0.273\textwidth}{0.398\textwidth}{\small (\fff) $\tau_{zz}$, 3D-TFM}%
\mylab{-0.860\textwidth}{0.233\textwidth}{\small (\ggg) $\tau_{xz}$, 2D-TFM \cite{delalamo2007}}%
\mylab{-0.565\textwidth}{0.233\textwidth}{\small (\hhh) $\tau_{yz}$, 2D-TFM \cite{delalamo2007}}%
\mylab{-0.273\textwidth}{0.233\textwidth}{\small (\iii) $\tau_{zz}$, 2D-TFM \cite{delalamo2007}}%
\mylab{-0.860\textwidth}{0.068\textwidth}{\small (\jjj) $\tau_{xz}$, 2D-TFM \cite{trepat2009physical}}%
\mylab{-0.565\textwidth}{0.068\textwidth}{\small (\kkk) $\tau_{yz}$, 2D-TFM \cite{trepat2009physical}}%
\mylab{-0.273\textwidth}{0.068\textwidth}{\small (\lll) $\tau_{zz}$, 2D-TFM \cite{trepat2009physical}}%
\caption{ Side by side comparison of 3D Fourier TFM versus
previous 2D methods \cite{delalamo2007,trepat2009physical} for a synthetic
deformation field representative of the deformation pattern exerted by
migrating amoeboid cells (see figure \ref{fig:examplecell}).  The Poisson's ratio
is $\sigma = 0.3$ and the substratum thickness, $h = 2 \Delta$, is equal to the
length of the ``synthetic cell''.
The plots in the top row show the synthetic deformation field in the $x$
direction (eq. \ref{eq:usyn}, panel \aaa),  $y$ direction (zero, panel \bbb)
and $z$ direction (eq. \ref{eq:wsyn}, panel \ccc). 
The second row shows the traction stresses calculated from the displacements in
panels (\aaa)-(\ccc) by 3D Fourier TFM.  (\ddd), $\tau_{xz}$;
(\eee), $\tau_{yz}$; (\fff), $\tau_{zz}$. 
The third row shows the traction stresses calculated from the displacements in
panels (\aaa)-(\ccc) by 2D Fourier TFM under the assumption of
zero normal displacements on the substratum's surface (\ie $w(z=h)=0$ as in ref.
\cite{trepat2009physical}).  (\ggg), $\tau_{xz}$; (\hhh), $\tau_{yz}$; (\iii),
$\tau_{zz}$. 
The last row shows the traction stresses calculated from the displacements in
panels (\aaa)-(\ccc) by 2D Fourier TFM under the assumption of
zero normal stresses on the substratum's surface (\ie $\tau_{zz}(z=h)=0$ as in
ref.  \cite{delalamo2007}).  (\jjj), $\tau_{xz}$; (\kkk), $\tau_{yz}$; (\lll),
$\tau_{zz}$. 
}
\label{fig:synthe03gau}
\end{center}
\end{figure}

\begin{figure}[!ht]
\begin{center}
\includegraphics[width=0.5\textwidth]{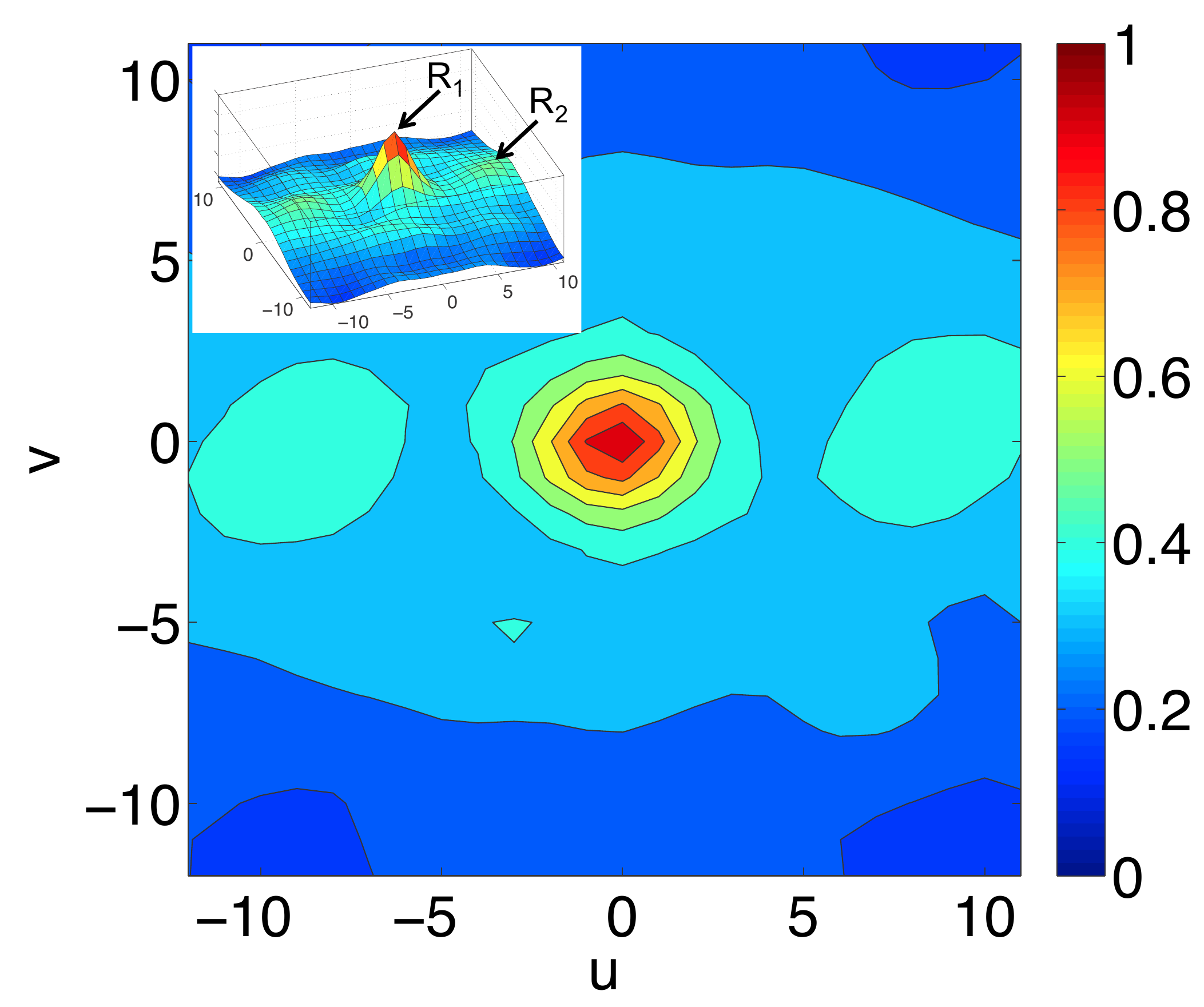}
\\
\hspace{0.1ex}
\includegraphics[width=0.5\textwidth]{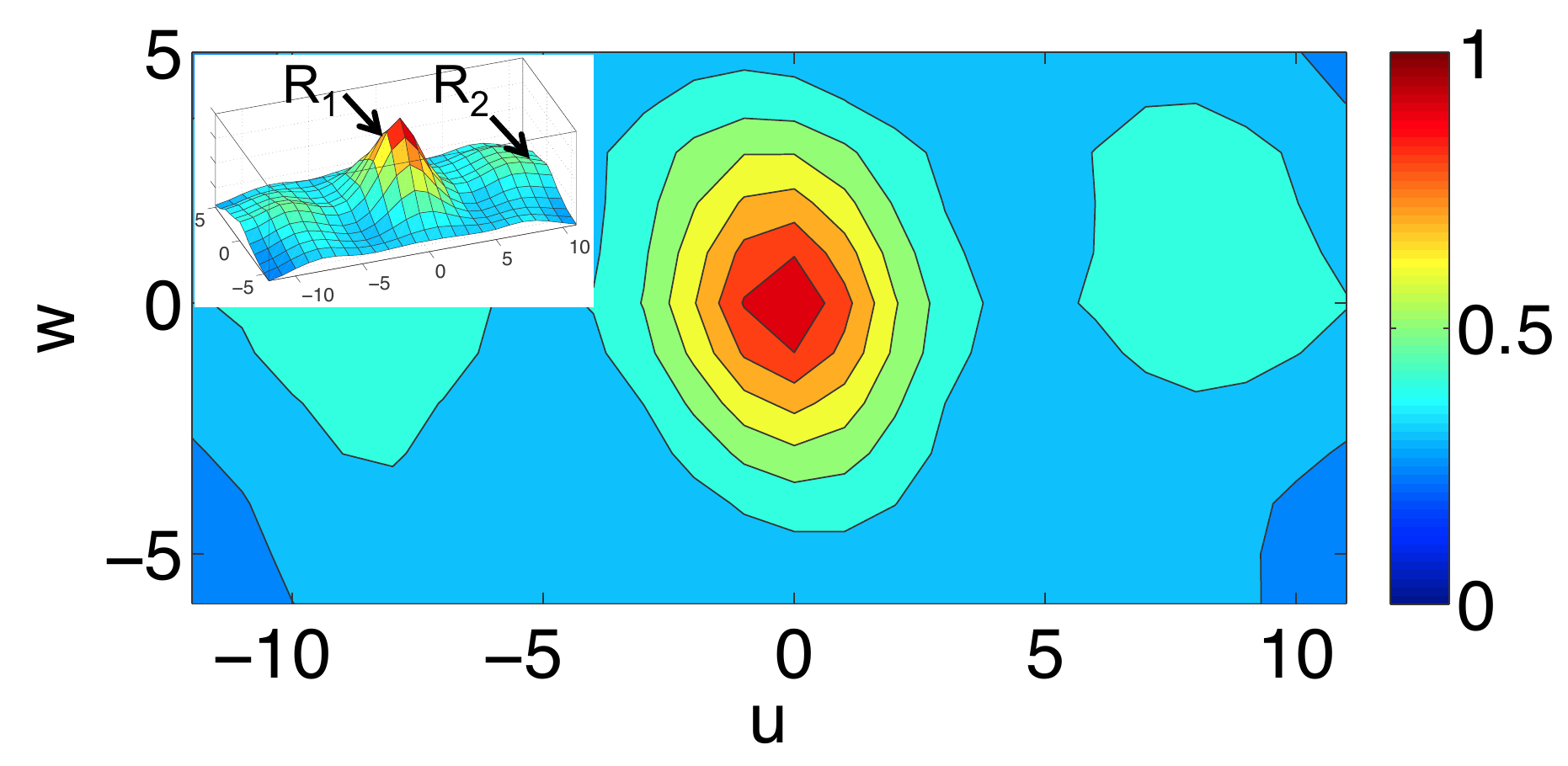}
\mylab{-0.53\textwidth}{0.63\textwidth}{(\aaa)}%
\mylab{-0.53\textwidth}{0.22\textwidth}{(\bbb)}%
\caption{ 
Example of the three-dimensional cross-correlation of fluorescence intensity,
$R(u,v,w)$, for a pair of interrogation boxes of size $24 \times 24 \times 12$
in the $x$, $y$ and $z$ directions.  The three-dimensional location of the peak
of the cross-correlation yields the relative displacement between the two
interrogation boxes.  The signal-to-noise ratio in this example, $s2n = 2.22$,
is determined by the ratio of the maximum value of the cross-correlation
($R_1=1$) to the second highest local maximum ($R_2 = 0.45$).  
(\aaa), Contour map of a two-dimensional section of the cross-correlation for
zero displacement in the $z$ direction, $R(u,v,w=0)$.
(\bbb), Contour map of a two-dimensional section of the cross-correlation for
zero displacement in $y$ direction, $R(u,v=0,w)$.
The insets in both panels are height maps of each two-dimensional section of
$R(u,v,w)$.
}
\label{fig:s2n}
\end{center}
\end{figure}

\begin{figure}[!ht]
\begin{center}
\includegraphics[width=0.6\textwidth]{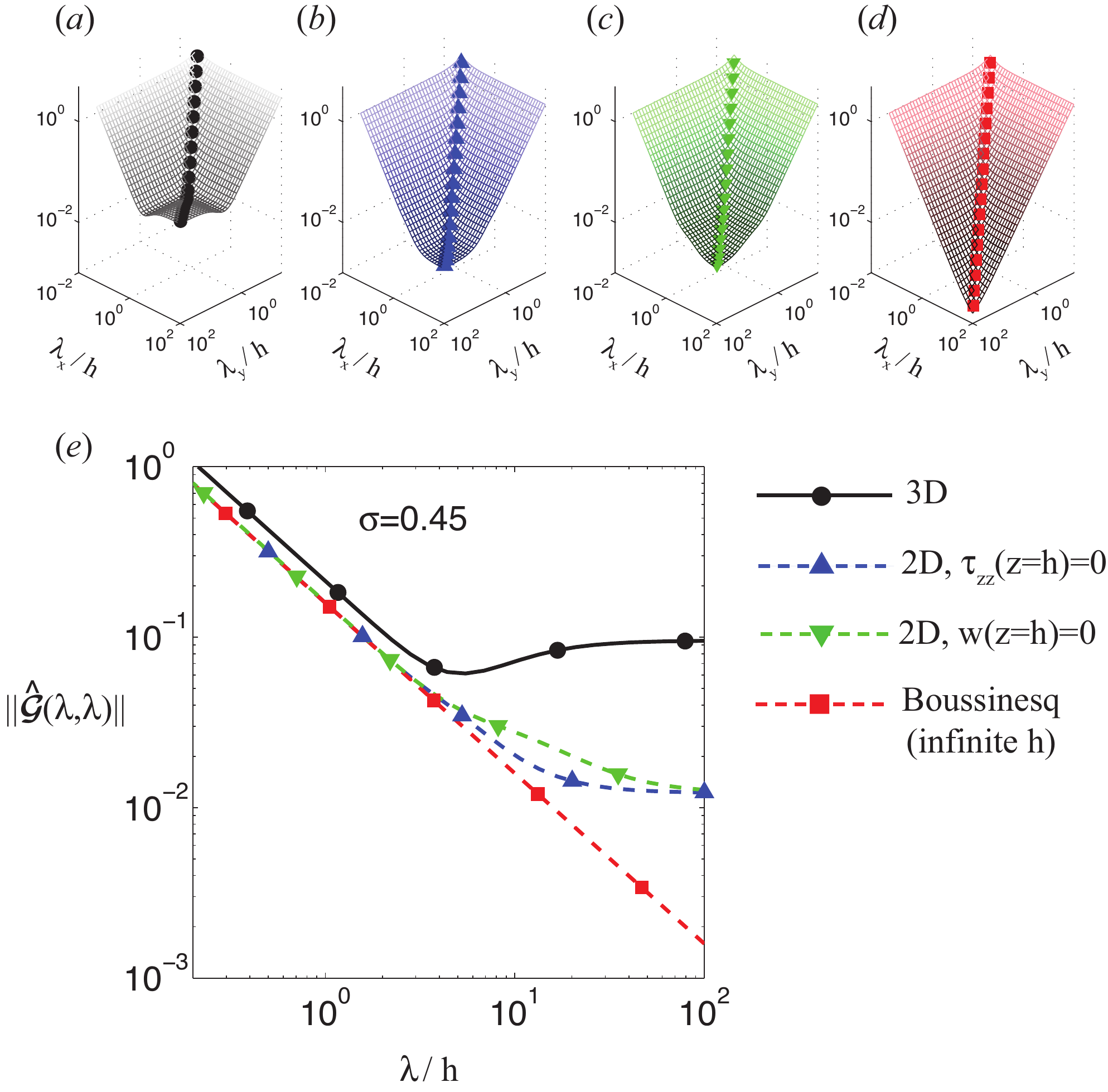}
\caption{Fr\"obenius norm of the Green's function used by different TFM
methods, $||{\bf \mathcal{\widehat G}}||$ (eq. \ref{eq:norma}), for a value of
the Poisson's ratio $\sigma = 0.45$.  The four panels in the top row (\aaa--\ddd)
show surface plots of $||{\bf \mathcal{\widehat G}}||$ as a function of the
horizontal wavelengths of the deformation field $(\lambda_x,\lambda_y)$.  
\abol, present 3D TFM method; 
\bbol, 2D TFM under the assumption of zero normal stresses on the substratum's
surface (\ie $\tau_{zz}(z=h)=0$ as in ref.  \cite{delalamo2007}); 
\cbol, 2D TFM under the assumption of zero normal displacements on the
substratum's surface (\ie $w(z=h)=0$ as in ref.  \cite{trepat2009physical});
\dbol, Boussinesq's traction cytrometry assuming infinitely-thick substratum
(as in refs.  \cite{dembo1996,butler2002}).  The symbol curves in these plots
indicate the sections of $||{\bf \mathcal{\widehat G}}||$ represented in panel
(\eee).  \ebol,  $||{\bf \mathcal{\widehat G}}||$ along the line $\lambda =
\lambda_x = \lambda_y$ from different TFM methods, represented as a function of
$\lambda / h$.
}
\label{fig:kerncomp2}
\end{center}
\end{figure}

\end{document}